\documentclass[numberedappendix]{emulateapj}

\usepackage{color}
\usepackage{rotating}

\begin{document}

\title{The Difference Imaging Pipeline for the Transient Search  \\ in the Dark Energy Survey}

\author{
R.~Kessler\altaffilmark{1,2},
J.~Marriner\altaffilmark{3},
M.~Childress\altaffilmark{4,5},
R.~Covarrubias\altaffilmark{6},
C.~B.~D'Andrea\altaffilmark{7},
D.~A.~Finley\altaffilmark{3},
J.~Fischer\altaffilmark{8},
R.~J.~Foley\altaffilmark{9,10},
D.~Goldstein\altaffilmark{11,12},
R.~R.~Gupta\altaffilmark{13},
K.~Kuehn\altaffilmark{14},
M.~Marcha\altaffilmark{15},
R.~C.~Nichol\altaffilmark{7},
A.~Papadopoulos\altaffilmark{7},
M.~Sako\altaffilmark{8},
D.~Scolnic\altaffilmark{1},
M.~Smith\altaffilmark{16},
M.~Sullivan\altaffilmark{16},
W.~Wester\altaffilmark{3},
F.~Yuan\altaffilmark{4,5},
T.~Abbott\altaffilmark{17},
F.~B.~Abdalla\altaffilmark{15,18},
S.~Allam\altaffilmark{3},
A.~Benoit-L{\'e}vy\altaffilmark{15},
G.~M.~Bernstein\altaffilmark{8},
E.~Bertin\altaffilmark{19,20},
D.~Brooks\altaffilmark{15},
A.~Carnero~Rosell\altaffilmark{21,22},
M.~Carrasco~Kind\altaffilmark{9,6},
F.~J.~Castander\altaffilmark{23},
M.~Crocce\altaffilmark{23},
L.~N.~da Costa\altaffilmark{21,22},
S.~Desai\altaffilmark{24,25},
H.~T.~Diehl\altaffilmark{3},
T.~F.~Eifler\altaffilmark{8,26},
A.~Fausti Neto\altaffilmark{21},
B.~Flaugher\altaffilmark{3},
J.~Frieman\altaffilmark{3,1,2},
D.~W.~Gerdes\altaffilmark{27},
D.~Gruen\altaffilmark{28,29},
R.~A.~Gruendl\altaffilmark{9,6},
K.~Honscheid\altaffilmark{30,31},
D.~J.~James\altaffilmark{17},
N.~Kuropatkin\altaffilmark{3},
T.~S.~Li\altaffilmark{32},
M.~A.~G.~Maia\altaffilmark{21,22},
J.~L.~Marshall\altaffilmark{32},
P.~Martini\altaffilmark{30,33},
C.J.~Miller\altaffilmark{34,27},
R.~Miquel\altaffilmark{35},
B.~Nord\altaffilmark{3},
R.~Ogando\altaffilmark{21,22},
A.~A.~Plazas\altaffilmark{26},
K.~Reil\altaffilmark{36},
A.~K.~Romer\altaffilmark{37},
A.~Roodman\altaffilmark{38,36},
E.~Sanchez\altaffilmark{39},
I.~Sevilla-Noarbe\altaffilmark{39,9},
R.~C.~Smith\altaffilmark{17},
M.~Soares-Santos\altaffilmark{3},
F.~Sobreira\altaffilmark{3,21},
G.~Tarle\altaffilmark{27},
J.~Thaler\altaffilmark{10},
R.~C.~Thomas\altaffilmark{12},
D.~Tucker\altaffilmark{3},
A.~R.~Walker\altaffilmark{17}
\\ \vspace{0.2cm} (The DES Collaboration) \\
}
 
\altaffiltext{1}{Kavli Institute for Cosmological Physics, University of Chicago, Chicago, IL 60637, USA}
\altaffiltext{2}{Department of Astronomy and Astrophysics, University of Chicago, 5640 South Ellis Avenue, Chicago, IL 60637, USA}
\altaffiltext{3}{Fermi National Accelerator Laboratory, P.O. Box 500, Batavia, IL 60510, USA}
\altaffiltext{4}{ARC Centre of Excellence for All-sky Astrophysics (CAASTRO), Australian National University, Canberra, ACT 2611, Australia.}
\altaffiltext{5}{The Research School of Astronomy and Astrophysics, Australian National University, ACT 2601, Australia}
\altaffiltext{6}{National Center for Supercomputing Applications, 1205 West Clark St., Urbana, IL 61801, USA}
\altaffiltext{7}{Institute of Cosmology \& Gravitation, University of Portsmouth, Portsmouth, PO1 3FX, UK}
\altaffiltext{8}{Department of Physics and Astronomy, University of Pennsylvania, Philadelphia, PA 19104, USA}
\altaffiltext{9}{Department of Astronomy, University of Illinois,1002 W. Green Street, Urbana, IL 61801, USA}
\altaffiltext{10}{Department of Physics, University of Illinois, 1110 W. Green St., Urbana, IL 61801, USA}
\altaffiltext{11}{Department of Astronomy, University of California, Berkeley,  501 Campbell Hall, Berkeley, CA 94720, USA}
\altaffiltext{12}{Lawrence Berkeley National Laboratory, 1 Cyclotron Road, Berkeley, CA 94720, USA}
\altaffiltext{13}{Argonne National Laboratory, 9700 South Cass Avenue, Lemont, IL 60439, USA}
\altaffiltext{14}{Australian Astronomical Observatory, North Ryde, NSW 2113, Australia}
\altaffiltext{15}{Department of Physics \& Astronomy, University College London, Gower Street, London, WC1E 6BT, UK}
\altaffiltext{16}{School of Physics and Astronomy, University of Southampton,  Southampton, SO17 1BJ, UK}
\altaffiltext{17}{Cerro Tololo Inter-American Observatory, National Optical Astronomy Observatory, Casilla 603, La Serena, Chile}
\altaffiltext{18}{Department of Physics and Electronics, Rhodes University, PO Box 94, Grahamstown, 6140, South Africa}
\altaffiltext{19}{Sorbonne Universit\'es, UPMC Univ Paris 06, UMR 7095, Institut d'Astrophysique de Paris, F-75014, Paris, France}
\altaffiltext{20}{CNRS, UMR 7095, Institut d'Astrophysique de Paris, F-75014, Paris, France}
\altaffiltext{21}{Laborat\'orio Interinstitucional de e-Astronomia - LIneA, Rua Gal. Jos\'e Cristino 77, Rio de Janeiro, RJ - 20921-400, Brazil}
\altaffiltext{22}{Observat\'orio Nacional, Rua Gal. Jos\'e Cristino 77, Rio de Janeiro, RJ - 20921-400, Brazil}
\altaffiltext{23}{Institut de Ci\`encies de l'Espai, IEEC-CSIC, Campus UAB, Facultat de Ci\`encies, Torre C5 par-2, 08193 Bellaterra, Barcelona, Spain}
\altaffiltext{24}{Department of Physics, Ludwig-Maximilians-Universitaet, Scheinerstr. 1, 81679 Muenchen, Germany}
\altaffiltext{25}{Excellence Cluster Universe, Boltzmannstr.\ 2, 85748 Garching, Germany}
\altaffiltext{26}{Jet Propulsion Laboratory, California Institute of Technology, 4800 Oak Grove Dr., Pasadena, CA 91109, USA}
\altaffiltext{27}{Department of Physics, University of Michigan, Ann Arbor, MI 48109, USA}
\altaffiltext{28}{Max Planck Institute for Extraterrestrial Physics, Giessenbachstrasse, 85748 Garching, Germany}
\altaffiltext{29}{Universit\"ats-Sternwarte, Fakult\"at f\"ur Physik, Ludwig-Maximilians Universit\"at M\"unchen, Scheinerstr. 1, 81679 M\"unchen, Germany}
\altaffiltext{30}{Center for Cosmology and Astro-Particle Physics, The Ohio State University, Columbus, OH 43210, USA}
\altaffiltext{31}{Department of Physics, The Ohio State University, Columbus, OH 43210, USA}
\altaffiltext{32}{George P. and Cynthia Woods Mitchell Institute for Fundamental Physics and Astronomy, and Department of Physics and Astronomy, Texas A\&M University, College Station, TX 77843,  USA}
\altaffiltext{33}{Department of Astronomy, The Ohio State University, Columbus, OH 43210, USA}
\altaffiltext{34}{Department of Astronomy, University of Michigan, Ann Arbor, MI 48109, USA}
\altaffiltext{35}{Institut de F\'{\i}sica d'Altes Energies, Universitat Aut\`onoma de Barcelona, E-08193 Bellaterra, Barcelona, Spain}
\altaffiltext{36}{SLAC National Accelerator Laboratory, Menlo Park, CA 94025, USA}
\altaffiltext{37}{Department of Physics and Astronomy, Pevensey Building, University of Sussex, Brighton, BN1 9QH, UK}
\altaffiltext{38}{Kavli Institute for Particle Astrophysics \& Cosmology, P.O. Box 2450, Stanford University, Stanford, CA 94305, USA}
\altaffiltext{39}{Centro de Investigaciones Energ\'eticas, Medioambientales y Tecnol\'ogicas (CIEMAT), Madrid, Spain}

\newcommand{\SNANA}{{\tt SNANA}}
\newcommand{\PSNID}{{\tt PSNID}}
\newcommand{\SALTII}{{\sc SALT-II}}
\newcommand{\SDSS}{SDSS-II}
\newcommand{\PS}{Pan-STARRS1}
\newcommand{\Diff}{{\tt DiffImg}}
\newcommand{\hotPants}{{\tt hotPants}}
\newcommand{\autoScan}{{\tt autoScan}}
\newcommand{\DESSN}{DES-SN}
\newcommand{\SBa}{SB anomaly}

\newcommand{\sex}{{\tt SExtractor}}
\newcommand{\scamp}{{\tt SCAMP}}
\newcommand{\swarp}{{\tt SWarp}}
\newcommand{\psfex}{{\tt PSFEx}}

\newcommand{\DES}{Dark Energy Survey}
\newcommand{\Spec}{Spectroscopic}
\newcommand{\spec}{spectroscopic}
\newcommand{\specy}{spectroscopically}
\newcommand{\obs}{observation}
\newcommand{\obss}{observations}
\newcommand{\eff}{efficiency}
\newcommand{\ineff}{inefficiency}
\newcommand{\effs}{efficiencies}
\newcommand{\unc}{uncertainty}
\newcommand{\uncs}{uncertainties}
\newcommand{\cand}{candidate}
\newcommand{\cands}{candidates}

\newcommand{\photoz}{photo-$z$}
\newcommand{\zphot}{z_{\rm phot}}
\newcommand{\OL}{\Omega_{\Lambda}}
\newcommand{\OM}{\Omega_{\rm M}}
\newcommand{\effCand}{{\mathcal E}_{\rm cand}}
\newcommand{\effPSNID}{{\mathcal E}_{\rm PSNID}}
\newcommand{\effSNR}{{\epsilon}_{\rm S/N}}
\newcommand{\sigSIM}{\sigma_{\rm SIM}}
\newcommand{\sigSB}{\sigma_{\rm SB}}
\newcommand{\SNRcalc}{{\rm SNR}_{\rm calc}}
\newcommand{\SNRdiff}{{\rm SNR}_{\rm\tt DiffImg}}
\newcommand{\SNRvingt}{ \overline{{\rm S/N}}_{\rm mag20}}
\newcommand{\Ncand}{N_{\rm cand}}
\newcommand{\Tobs}{T_{\rm obs}}
\newcommand{\Trest}{T_{\rm rest}}
\newcommand{\txy}{t_{x,y}}
\newcommand{\tcxy}{t_{x,y}^{\prime}}
\newcommand{\mhalf}{m_{{\rm eff}=1/2}}
\newcommand{\Pfit}{P_{\rm fit}}
\newcommand{\chisqSigma}{\chi^2_{\sigma}}
\newcommand{\DF}{\Delta F}
\newcommand{\sigF}{\sigma_F}
\newcommand{\mSB}{m_{\rm SB}}
\newcommand{\rmsD}{{\rm RMS}_{\Delta}}
\newcommand{\restB}{\tilde{m}_B}
\newcommand{\degsq}{deg$^2$}
\newcommand{\SKYmin}{\sigma_{\rm SKY}^{\rm min}}
\newcommand{\sigSKY}{\sigma_{\rm SKY}}
\newcommand{\sigSKYideal}{\sigma_{\rm SKY}{\rm (ideal)}}
\newcommand{\Ntemplate}{N_{\rm template}}
\newcommand{\lbar}{\langle\lambda\rangle}
\newcommand{\Texpose}{T_{\rm expose}}
\newcommand{\texpose}{t_{\rm expose1}}
\newcommand{\Nepoch}{N_{\rm epoch}}
\newcommand{\Nexpose}{N_{\rm expose}}
\newcommand{\avgNepoch}{\langle{\Nepoch}\rangle}
\newcommand{\dlr}{d_{\rm LR}}
\newcommand{\Nvisit}{N_{\rm visit}}
\newcommand{\ndetect}{\bar{n}_{\rm detect}}

\newcommand{\MAXDATAVOL}{170}       
\newcommand{\FOV}{2.7}                              
\newcommand{\VETOAREAsqdeg}{0.63}    
\newcommand{\VETOAREApcnt}{2.4}         
\newcommand{\NRAWCANDTOT}{$1.2\times 10^5$}      
\newcommand{\NTMPCANDTOT}{$1.0\times 10^5$}      
\newcommand{\NSCICANDTOT}{7489}          
\newcommand{\autoScanRej}{13}              
\newcommand{\NFAKEPSNID}{7662}        
\newcommand{\zDeepEffHalf}{1.1}                 
\newcommand{\zShallowEffHalf}{0.7}                 
\newcommand{\NcandPerDeep}{1040}      
\newcommand{\NcandPerShallow}{680}      
\newcommand{\NSNCandMC}{$2000\pm 300$}   
\newcommand{\candFracSN}{27}                   
\newcommand{\candFracAGN}{30}                
\newcommand{\candFracSTAR}{10}               
\newcommand{\candFracCRAP}{30}                

\newcommand{\NDETECTg}{110}                      
\newcommand{\NDETECTz}{150}                      
\newcommand{\NDETECT}{130}                       

\email{kessler@kicp.uchicago.edu}
\submitted{Accepted by AJ,  2015 August 25}

\begin{abstract}
We describe the operation and performance of the difference imaging pipeline (\Diff) used to 
detect transients in deep images from the Dark Energy Survey Supernova program  (\DESSN) 
in its first observing season from  2013 August through  2014 February. \DESSN\ is a search for transients 
in which ten 3-\degsq\ fields are repeatedly observed in the $g,r,i,z$ passbands with a cadence 
of about 1 week. The observing strategy has been optimized to measure high-quality light curves 
and redshifts for thousands of Type Ia supernova (SN~Ia) with the goal of measuring dark energy 
parameters. The essential \Diff\ functions are to align each search image to a deep reference image,
do a pixel-by-pixel subtraction, and then examine the subtracted image for significant positive 
detections of point-source objects. The vast majority of detections are subtraction artifacts, but after 
selection requirements and image filtering with an automated scanning program, there are 
$\sim \NDETECT$ detections per \degsq\ per \obs\ in each band, of which only  $\sim 25$\% 
are artifacts. Of the $\sim 7500$  transients discovered by \DESSN\ in its first observing season,
each requiring a detection on at least two separate nights, 
Monte Carlo (MC) simulations predict that  $\candFracSN$\% are expected to be 
SNe~Ia or core-collapse SNe. Another $\sim \candFracCRAP$\% 
of the transients are artifacts in which a small number of \obss\ satisfy the selection criteria for a 
single-epoch detection. \Spec\ analysis shows that most of the remaining transients are AGN and 
variable stars. Fake SNe~Ia are overlaid onto the images to rigorously evaluate detection \effs\ 
and to understand the \Diff\ performance.  The \Diff\ \eff\ measured with fake SNe agrees well with 
expectations from a MC simulation that uses analytical calculations of the fluxes 
and their  \uncs. In our 8  ``shallow" fields with single-epoch 50\% completeness 
depth $\sim 23.5$, the SN~Ia  \eff\ falls to 1/2 at redshift $z\approx \zShallowEffHalf$; 
in our 2  ``deep" fields with mag-depth  $\sim 24.5$, the \eff\ falls to 1/2 at 
$z\approx \zDeepEffHalf$. A remaining performance issue is 
that the measured fluxes have additional scatter (beyond Poisson fluctuations) that increases with 
the host galaxy surface brightness at the transient location. This bright-galaxy issue has minimal 
impact on the SNe~Ia program,but it may lower the \eff\ for finding fainter transients on bright galaxies.
\end{abstract}

 \section{Introduction}
 \label{sec:intro}

The discovery of the accelerating expansion of the universe \citep{Riess98,Saul99} using 
Type Ia supernovae (SN Ia) has greatly motivated ever larger transient searches in broadband 
imaging surveys. The associated search pipelines have become increasingly complex in 
distributing enormous computing tasks needed to rapidly  find new transients for \spec\  \obss, 
and in processing a wide range of data quality.

A new era of  transient searches began in the early 2000s with ``rolling searches" in which the 
same telescope is used for discovering new objects and providing precise photometric 
measurements of the light curve in multiple passbands. To collect large SN~Ia samples for 
measuring cosmological parameters, the earliest rolling searches include 
the Supernova Legacy Survey (SNLS: \citealt{Astier2006,Perrett2010}),
ESSENCE \citep{Gajus2007}, and
the Sloan Digital Sky Survey-II (\SDSS: \citealt{Frieman2008,Sako2008}).
Each of these surveys discovered  many hundreds of
SNe~Ia, about half of which were \specy\ confirmed. 
The next generation of rolling searches includes the recently completed
\PS\  \citep{PS_2002},
the ongoing \DES\ (DES: \citealt{DESSN2012}), and
the Large Synoptic Survey Telescope (LSST:  \citealt{Ivezic2008,LSST_SciBook}),
expected to begin in the next decade.
Another advantage of these rolling searches is that there  is a complementary wide-area survey 
with the same instrument; this benefits the absolute calibration by including dithered exposures 
over the SN fields to inter calibrate the CCDs, regular \obss\ of standard-star fields, and measurements 
of the telescope and atmospheric transmission functions.

The goal of this paper is to describe the difference-imaging pipeline (\Diff) used to discover 
point-source transients in DES. We present detailed performance results of \Diff\ for single-epoch
detections, and for the redshift dependence of discovering and classifying SN~Ia light curves.
While the search strategy was  optimized to find SNe~Ia to build a Hubble diagram for measuring 
dark energy properties, \citep{DESSN2012}, \Diff\ does not depend on the transient type. 
In addition to SNe~Ia, our \Diff\ has found many other transient types including
core-collapse SNe (CC SNe),
super-luminous SNe (SLSNe: \citealt{Papa2015}),
active galactic nuclei (AGNs),
Kuiper belt objects (KBOs: \citealt{KBO2015}),
and a possible tidal disruption event \citep{Rees1988,TDE2015}.

The challenges for  \Diff\ are to 
produce a high quality subtracted image for each search image
by subtracting a deep coadded template,
reject a large number of non-astrophysical detections (artifacts) in the subtracted images,
develop a workflow to process each night of data in less than a day, 
and monitor the performance well enough to uncover subtle problems 
and to determine \effs\ and biases for  science analyses.
We use publicly available codes for the core routines needed to
determine an astrometric solution,
co-add exposures, 
measure the point-spread function (PSF),
align template and search images,
perform the subtractions,
and fit light curves to a series of SN templates for classification.
In addition to these existing codes, we have  developed new software tools for
automated scanning of subtracted images \citep[hereafter G15]{autoScan}, 
detailed monitoring based on artificial SNe overlaid on images,
and a workflow to distribute jobs on arbitrary computing platforms.

The essential monitoring element is to inject fake SNe~Ia onto galaxies 
in real images (hereafter called ``fakes''). 
The Supernova Cosmology Project used fakes to monitor the
\eff\ of human scanners in the real-time SN~Ia search \citep{Pain2002},
and also to measure the analysis \eff\ as part of the SN~Ia rate measurement.
Fakes were later used for real-time monitoring in the \SDSS\ Supernova search \citep{Dilday2008} 
to measure the \eff\  of the detection pipeline and human scanning.
The Nearby Supernova Factory moved stars on the image to serve as fake transients;
they monitored their single-epoch detection \eff\
and trained their machine learning method that was used to reject large numbers
of subtraction artifacts \citep{SNF2007}.
SNLS used fakes in an offline analysis \cite[hereafter P10]{Perrett2010}
to measure their \effs\ and selection biases that impact the Hubble diagram.
In \DESSN\  the fakes are used to 
(1) monitor the detection depth,
(2) monitor the single-epoch detection \eff\ 
    from the fraction of fakes that are detected,
(3) monitor the \eff\ for multiple detections that are required for \spec\ targeting and science analysis,
(4) train the automated scanning software (G15), and
(5) characterize the \Diff\ performance for a fast Monte Carlo (MC) simulation.

Compared to the use of fakes in previous surveys, an improvement in \DESSN\ 
is that the ideal \eff\ is predicted from the signal-to-noise ratio (S/N)
of the flux measurements, and thus 
the \Diff\ performance can be rigorously evaluated by comparing the
predicted and fake \effs.
This prediction, as a function of fake SN redshift, 
comes from a fast MC simulation that computes 
realistic light curves without using images or pixels. 
The fast MC simulation analytically computes the light curve fluxes and their \uncs\ 
using input from observed conditions, 
and also from key \Diff\ properties derived from the fakes.
The agreement (or lack of) between the predicted and measured
\eff\ provides a robust measure of the \Diff\ performance.

There is another practical motivation for using a fast MC simulation to 
validate the point-source \Diff\ \eff\ with fakes. 
Typical science analyses require large SN simulations that are repeated 
many times for development, evaluation of systematic \uncs, 
and estimates of contamination from CC SNe.
Ideally, such simulations would be similar to the fakes in which calculated 
light curves are overlaid on CCD images and processed with  \Diff.
The CPU resources for so many image-based simulations,  however, 
would be quite enormous.
On the other hand, the fast MC simulation in \SNANA\   \citep{SNANA} 
can generate close to $10^2$ light curves per second on a single core,
which is five orders of magnitude faster than the ideal image-based simulation. 
Our goal, therefore, is to use a single realization of fakes to characterize 
the \Diff\  performance for the fast MC simulation; the fast MC can then be 
used to rapidly generate samples of point-source transients  
with the same \effs\ and \uncs\ as an image-based simulation.
Although only one transient type (SN~Ia) is used to generate fakes for
image overlays and \Diff\ processing, the resulting fast MC simulation 
can in principle be used for any SN type, 
and more generally for any point-source transient.

The outline of this paper is as follows.
An overview of DES and the transient search is given in \S\ref{sec:survey},
and  \Diff\ is described in \S\ref{sec:diff}.
The monitoring of single-epoch detections is given in \S\ref{sec:monObs},
including the single-epoch magnitude depths, data quality evaluation,
\eff\ vs. S/N, and the anomalous scatter of flux measurements
for objects on bright galaxies.
The \eff\ of multiple detections required for a transient
is described in \S\ref{sec:monCand},
including the discovery \eff\ and the classification \eff.
In \S\ref{sec:datamc} we compare the simulation to data
in a preliminary photometric analysis. 
Comparisons with SNLS and \Diff\ limitations are discussed in  \S\ref{sec:discuss},
and we conclude in \S\ref{sec:fin}.

\section{Overview of the Dark Energy Survey and Transient Search}
\label{sec:survey}

The \DES\ includes a wide-area 5000~\degsq\ optical survey in the 
southern celestial hemisphere
and a dedicated transient search over 27~\degsq, 
both using the Dark Energy Camera
(DECam: \citealt{DECAM2015}).
DECam is mounted on the Blanco 4-m telescope at the 
Cerro Tololo Inter-American Observatory (CTIO) and the data are processed by the 
DES data management system \citep{DESDM2011,Mohr2012,Desai2012} at the
National Center for Supercomputing Applications  (NCSA).
The 570 Megapixel DECam has a 3 \degsq\ field of view and is composed
of 62  science-image CCDs, each with 2k $\times$ 4k pixels,
and 8 CCDs for guiding.
After accounting for CCD gaps and two non-functioning
CCDs, the active field of view is \FOV\ \degsq.

The transient search is performed in 10 ``SN fields'' (27 \degsq) 
that are repeatedly observed in the $g,r,i,z$ passbands.
We refer to this part of the survey as \DESSN.
Eight of these fields are observed with few-minute exposure times
and are referred to as ``shallow'' fields;
the remaining two ``deep'' fields are observed much longer (Table~\ref{tb:bands}).
Defining the AB magnitude-depth as the mag where the \Diff\  single-epoch detection
\eff\ has fallen to 50\%, the shallow and deep field depths are
$\sim 23.5$ and $\sim 24.5$, respectively, 
and the depth in each band is the same.
The SN portion of the DES observing strategy is that the wide-area survey 
transitions to observing SN fields 
when the seeing is above  $1.1\arcsec$,
and, in addition,  any SN field (in any band) which has not been observed 
for 7~days is scheduled with the highest observing priority
regardless of the seeing. This 7-day trigger typically results in better
data quality compared to the $1.1\arcsec$ trigger.
For each SN field, the pointing at a repeat visit is the same 
to within a  few arcseconds.
Additional dithered \obss\ from the wide-area survey are used for the
inter-calibration of the CCDs.

On a given night, the number of consecutive exposures ($\Nexpose$) varies
with band and field as shown in Table~\ref{tb:bands}.
$\Nexpose=1$ for the shallow $g,r,i$ bands where the sky level
is well below saturation. In the deep fields (and shallow $z$ band),
$\Nexpose >1$ to limit the sky level to be well below saturation
within each exposure.
For a shallow field, each observing block is scheduled for all four bands
and takes $\sim 20$~minutes with overhead. 
For a deep field, observing all exposures in each band takes about 2~hours,
and would be difficult to schedule such a long block within the constraints of
the global DES observing strategy. 
Each deep-field band is therefore scheduled independently; 
the total exposure time per epoch ($\Texpose$)  is $10$~minutes in the $g$ band,
and more than an hour in the $z$ band (Table~\ref{tb:bands}).

\begin{table}[h!]
\caption{  Exposure Summary of  \DESSN\ Fields.  }  
\begin{center}  \begin{tabular}{| l  | c | c  | c | c |}    \tableline  
    &          &   Central $\lambda$     &   $\Texpose$ (sec)\tablenotemark{a}  & Template \\
   &  Band     &   (\AA)                      &  per Epoch                      & $\avgNepoch$\tablenotemark{b}    \\
      \hline\hline 
 Shallow  &  $g$     &   4830          &  $1\times 175 = 175$   &  8.0 \\
               &   $r$     &   6430          &  $1\times 150 = 175$   &  8.5 \\
               &   $i$     &   7830          &  $1\times 200 = 200$   &  9.3 \\
               &   $z$    &   9180          &  $2\times 200 = 400$   &  9.3 \\
    \hline
 Deep   &    $g$     &   4830          &  $~3\times 200 = ~600$   &  5.5 \\
             &    $r$     &   6430          &  $~3\times 400 = 1200$   &  7.5 \\
             &    $i$      &   7830          &  $~5\times 360 = 1440$   &  9.3 \\
             &    $z$     &   9180          &  $11\times 330 = 3630$   &  8.3 \\
\tableline  \end{tabular} 
 \tablenotetext{1}{
  $\Nexpose \times \texpose = \Texpose$.
  }
 \tablenotetext{2}{
    Averaged over fields. 
    The average template exposure time is $\avgNepoch \times \Texpose$.
  }
\end{center}   \label{tb:bands} \end{table}

The ten SN fields are divided into four groups of adjacent fields:
three  C fields that overlap the Chandra deep fields,
three  X fields that overlap the XMM-LSS fields,
two S fields that overlap SDSS stripe 82, and 
two E fields that overlap the ELAIS S1 field.  
The field locations were chosen based on 
(1) visibility from CTIO,
(2) visibility from telescopes in the northern hemisphere to perform follow-up 
  spectroscopy of live targets,
(3) galactic extinction,
(4)  avoiding overlap with extremely bright stars, 
(5) overlap with pre-existing galaxy catalogs and calibration.
A summary of each field and its location is given in Table~\ref{tb:fields}.
The maximum nightly data volume from observing all ten fields
is \MAXDATAVOL~GB, corresponding to just over 5000 CCD images.
The average data volume in a typical night corresponds to a few fields.
In addition to the SN field \obss, \DESSN\ makes use
of the extensive calibration data (\S\ref{sss:calib}) taken as part
of survey operations.

\begin{table}[h!]
\caption{ \DESSN\  field names and locations  }  
\begin{center} \begin{tabular}{ | l  |  l rr | c | } \tableline  
                  &    Deep or  &  \multicolumn{2}{c | }{Field Center (deg): } &  
                           $\Nvisit$\tablenotemark{a}  \\
       Field    &    Shallow   &   R.A. &    Decl.       &    $g/r/i/z$  \\
   \hline\hline 
     C1  &  shallow  & 54.2743  &  $-27.1116$   &  29/30/30/30  \\
     C2  &  shallow  & 54.2743  &  $-29.0884$   & 28/28/27/28  \\
     C3  &  deep       & 52.6484  &  $-28.1000$  &  25/23/28/27  \\
    X1  & shallow   & 34.4757   &  $-4.9295$  &   26/27/27/27  \\
    X2  & shallow   & 35.6645   &  $-6.4121$  &   26/26/25/24  \\
    X3  & deep        & 36.4500    &  $-4.6000$  &   21/20/22/24  \\
   S1  & shallow & 42.8200    &  0.0000      &   29/29/28/28  \\
   S2  & shallow & 41.1944    &  $-0.9884$ &   27/28/28/28  \\
     E1  & shallow & 7.8744  &  $-43.0096$  &   27/26/27/26  \\
    E2   & shallow & 9.5000  &  $-43.9980$  &   26/26/26/27  \\
\tableline  \end{tabular} \end{center}   
\tablenotetext{1}{Number of single-epoch visits to each field in each band, in Y1.}
\label{tb:fields} \end{table}

Science Verification (SV) took place 2012 November  through 2013 January,
with the goal of ensuring that the DECam performance meets
the DES science requirements. During the beginning of SV, 
the SN fields were observed to obtain initial calibrations
and to build templates. The latter part of SV was used to test \Diff.
Nominal survey operations began in the Fall of 2013. The first season
(2013 August to  2014 February) is referred to as Y1, and the second season
(2014 August to  2015 February) is referred to as Y2.

\bigskip
\section{The Difference-Imaging Pipeline}
\label{sec:diff}

All images taken with DECam at CTIO are  transferred to NCSA and run through 
the detrending process to produce images suitable for higher level analyses.
For each exposure, all CCDs on the focal plane are processed as a single unit 
where bad pixels are masked and corrections are applied for 
bias, 
flat-field 
illumination,
pupil ghost, 
crosstalk,
linearity, and overscan.
More details are given in \cite{Mohr2012,Desai2012} and references within.
The detrending process is virtually identical for the SN fields and the wide-area survey,
and it is  similar to a community pipeline used to process DECam data for
non-DES observers.

\begin{figure*}  
\centering
\epsscale{1.2} 
\plotone{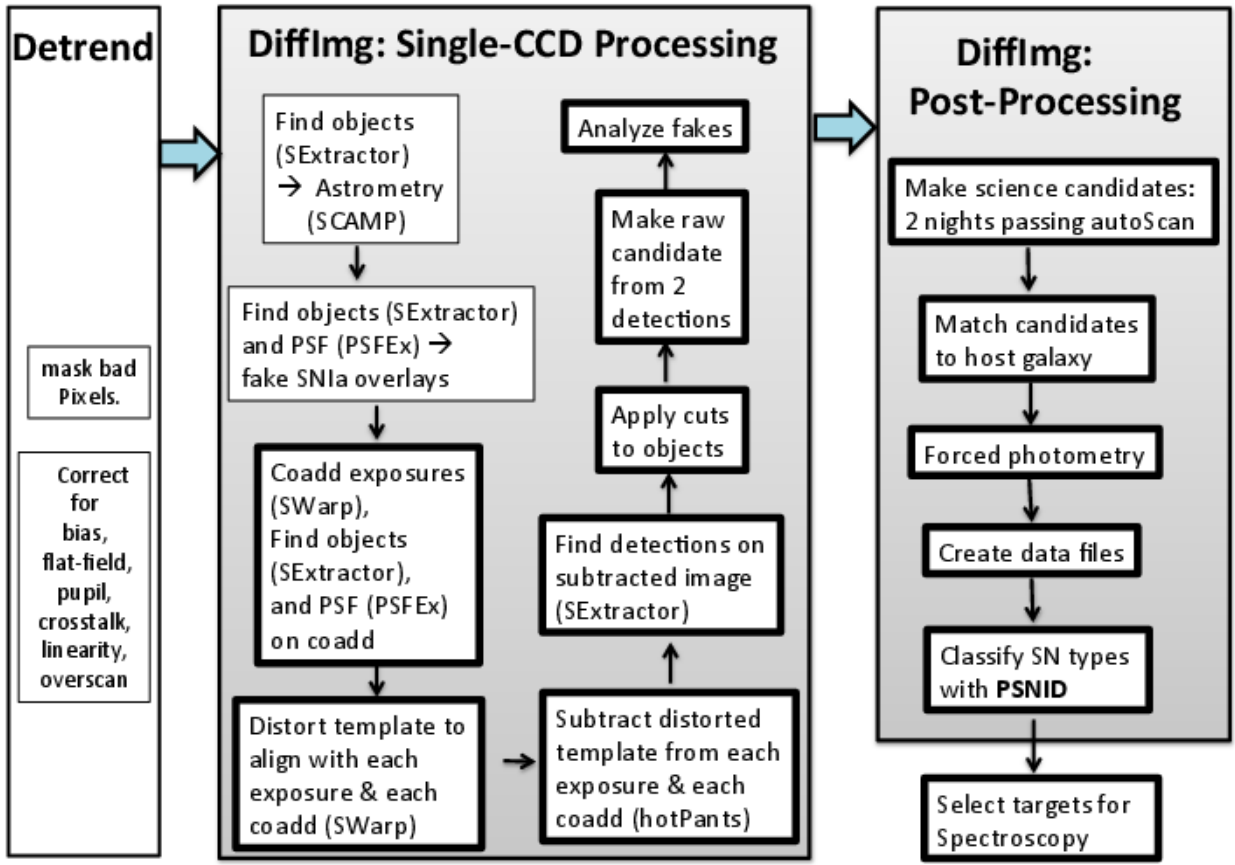}
\vspace{-1.5cm}
  \caption{
   Schematic overview of the  SN-field processing:
   detrending (left panel) and \Diff\ (middel+right panels).
   The thin-lined boxes refer to operations on individual exposures;
   the thick-lined boxes refer to operations on coadds.
   Astromatic.net codes are shown in parentheses.
  }
  \label{fig:diffim_overview}
\end{figure*}

For the SN fields, \Diff\ is run after the detrending and a schematic overview is shown in
Fig.~\ref{fig:diffim_overview}.  In contrast to detrending,
\Diff\ is run independently for each CCD in order to simplify
the distribution of jobs among CPUs.
Many of our \Diff\ stages use publicly available 
Terapix/AstrOmatic
codes\footnote{http://www.astromatic.net}\citep{Tpix2002} including 
{\scamp} \citep{SCAMP2006}  for astrometry, 
{\sex} \citep{sex1996} to find objects,
{\psfex}  \citep{PSFEx2011}  to determine the position-dependent PSF, and
{\swarp} to sum individual exposures (to make ``coadds'') and to align template images
      to search images.
The sub-sections below describe \Diff\ in more detail.      

\subsection{Pre-survey Observations and Analysis}
\label{ssec:preSurevy}

\subsubsection{Calibration}
\label{sss:calib}

For the results presented here, the calibration\footnote{The global DES calibration 
plan is available at \\ http://des-docdb.fnal.gov:8080/cgi-bin/ShowDocument?docid\\=6584\&version=7}
is determined from data taken during SV.
The calibration is needed to determine magnitudes for detected transients,
which are used to select transients of appropriate brightness for \spec\ \obss.
The calibration is also used to convert the fake magnitudes into fluxes in CCD counts.

During nightly operations, DES typically observes a set of 3 standard star fields corresponding 
to low, intermediate and high airmass.  These \obss\ are done during evening twilight, 
and again during morning twilight \citep[G.Bernstein et al. 2015, in prep]{Tucker2007}.
These standard star fields are mostly in SDSS stripe 82, but supplemented with additional
fields, mostly at decl. $\approx -45^{\circ}$ to $-40^{\circ}$.  
The stars in these fields, which we refer to as secondary standard stars, 
have had their magnitudes transformed into the defined 
DES ``natural'' system in which the color terms are close to zero. For photometric nights, 
these well calibrated secondary standard stars are used to determine a nightly calibration
consisting of zero points (ZPs), atmospheric extinction coefficients, and color terms needed to 
transform the photometry from the individual DECam CCDs to the defined DES 
``natural'' system.   A set of standard stars within each of the ten SN fields,
referred to as ``tertiary standards," were calibrated from data taken during the
SV period under photometric conditions, using exposures centered on the SN fields plus 
additional dithered exposures from the wide-area survey; this resulted in typically 100-200 
well-calibrated tertiary standard stars per CCD-area in the SN fields (K.Wyatt et al. 2015, in prep).  
These tertiaries were used to calibrate the template images for \Diff\ (\S\ref{sss:templates}), 
and this calibration is transferred to each transient magnitude. The relative calibration between 
DES fields over large areas has been checked using the stellar locus regression method 
\citep{SLR2009,SLR2014}, where consistency of colors is verified at the 2\% level. 
The absolute calibration has been checked at the 2\% level using very short DES exposures 
on a handful of spectrophotometric standards measured by the Hubble Space Telescope.
While this early calibration  meets some of the DES requirements,
extensive efforts continue to significantly improve the calibration for analysis.

\subsubsection{Templates}
\label{sss:templates}

For Y1  we constructed deep coadded templates from the Y2 season,
while the calibration is from SV.
Starting with the image that has the lowest sky noise ($\SKYmin$), 
up to 10 epochs are selected with the smallest PSF 
that have sky noise less than $2.5\cdot\SKYmin$.
The average number of coadded epochs per band is shown in Table~\ref{tb:bands},
along with the total exposure time per epoch. 
In the deep-field $i$ band, for example, the templates include a CCD-average
of 9.3 epochs which corresponds to a total exposure of 3.7~hr.
With an average of 8 coadded epochs per template,  
the image-subtracted sky-noise  ($\sigSKY$) 
is only 6\% higher compared to using an ideal template with 
infinite S/N.\footnote{
$\sigSKY/\sigSKYideal \simeq \sqrt{(1+1/\Ntemplate)}$ where
$\Ntemplate$ is the number of coadded templates.} 

Calibrated tertiary standards (\S\ref{sss:calib}) are used to determine the ZP
for each exposure, and the pixel flux values in each exposure are re-scaled
to a common zero point, ${\rm ZP}=31.1928$.\footnote{ZP$= 25 + 2.5\log_{10}(300)$,
where 25 is a nominal ZP per second and 300~sec is a reference exposure time.}
The coadded templates are combined with a weighted average of each exposure,
and the weight within each CCD  is fixed to the inverse of the average sky-variance.

The astrometric alignment was done in two steps. 
First, the exposures were aligned to the USNO-B1 catalog \citep{USNO-B}
and then coadded to produce an intermediate set of templates in which the 
alignment is good to $\sim 100$~mas, or 0.4 pixel.
These intermediate templates were used to produce an internal DES catalog
based on \sex\ output.
Next, the exposures were re-aligned to this internal catalog,
resulting in $\sim 20$~mas (0.08 pixel) precision.
After this final astrometric alignment, the exposures are
coadded again to produce the final set of templates.

\bigskip\bigskip
\subsection{Single-CCD Processing}
\label{ssec:proc1ccd}

\subsubsection{Astrometry} \label{sss:astrom}

During SV and Y1, the astrometric solution was obtained for the entire focal 
plane in the detrending process, 
using the \scamp\ program and 
the UCAC-4 catalog \citep{UCAC-4} as an astrometric reference. 
While this worked well for the wide-area survey, there were sometimes 
very poor solutions in the SN fields leading to errors up to an arcsecond. 
We suspected that bright saturated objects contributed to this problem 
because of the longer exposure times in the SN fields compared to the 
wide-area survey.

After Y1, two astrometry updates were incorporated.
First, we switched to using  a fainter reference catalog in the SN fields, 
USNO-B \citep{USNO-B}.  
Second, rather than using \scamp\ to separately find an astrometric 
solution for the search and template images, we used the \scamp\ 
feature allowing a joint astrometric solution for the search and template images.
While the absolute astrometric precision of the USNO-B catalog (250 mas)
is worse than UCAC-4 (60 mas),
the second change ensures good astrometric alignment ($<30$ mas) 
between the search and template, which is critical for good subtractions.

These changes were not incorporated into the detrending process,
and  were instead added to \Diff. Since \Diff\ is designed for single-CCD 
processing, a {\scamp} solution is obtained separately for each CCD rather 
than over the focal plane. 
The astrometric changes worked significantly better, but a few percent  of the 
processed CCDs still suffered catastrophic failures in the astrometric solution. 
As a final refinement to eliminate these catastrophic solutions, 
we used our own DES data to construct a reference catalog
(\S\ref{sss:templates}).

\subsubsection{Overlaying Fakes onto Images} \label{sss:fakes}

In the next stage, two classes of fake point sources are overlaid on
the CCD  image. The first class consists of  four 20th mag fakes in each band
(hereafter called ``MAG20'' fakes) overlaid in random locations
away from masked regions. 
The resulting S/N from the \Diff\ flux measurements
is part of the data quality evaluation (\S\ref{sec:monObs}).

The second class of fakes, ``SN fakes,'' consists of  SN~Ia light curve fluxes 
overlaid onto the CCD image near real galaxies. 
The fake SN~Ia light curve magnitudes are generated by the 
\SNANA\ simulation \citep{SNANA},
and include true parent populations of stretch and color,
a realistic model of intrinsic scatter \citep{Guy2010,K13}, 
a redshift range from 0.1 to 1.4, and a galaxy location 
chosen randomly with a probability proportional to its surface brightness density.
All fake SN~Ia light curves are generated and stored prior to the start of the survey
in order to simplify the overlay software in \Diff.
The fake SN~Ia flux added to the image is determined by 
a ZP based on  the comparison of calibration star 
magnitudes with their fluxes recovered by \sex.
The SN flux is spread over nearby pixels using the PSF found by 
the program \psfex,
and the flux in each pixel is smeared by random Poisson noise.

\begin{figure}[h!]  \centering
\plotone{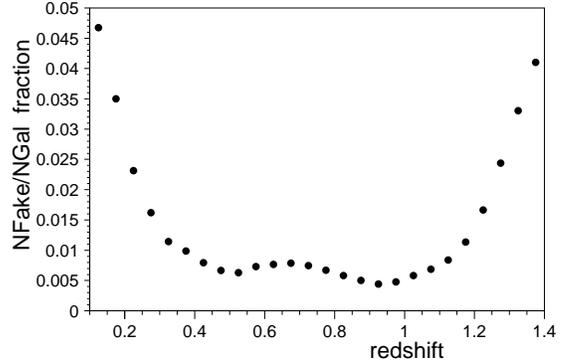}
  \caption{   
     As a function of redshift,  fraction of catalog  galaxies 
     in the SN fields  with an overlaid fake SN~Ia.
       }
  \label{fig:fakeGal}
\end{figure}

Ideally, fake SNe would be overlaid onto a duplicate set of images
so that images with and without fakes can be processed separately.
For \DESSN\ we did not prepare for this duplication,
and therefore care is taken to avoid consuming too many
galaxies with fake SNe 
that can overlap real transients and cause them to be undetected.
Fig.~\ref{fig:fakeGal} shows the fraction of catalog galaxies populated by fake 
SNe~Ia as a function of redshift; the redshift distribution has been sculpted
to ensure adequate low-redshift fakes for monitoring without
populating more than a few percent of the galaxies at the low and high
redshift ranges. 
At a given epoch, the average number of overlaid fakes per CCD is $\sim 20$.
Most of the overlaid fakes are far from peak or at high redshift, and
thus only about 1/3 of these are bright enough to be detected.

There are a few caveats regarding the selection of galaxies and the
placement of the fake.
First, simulated SNe~Ia are matched to a real galaxy based on the galaxy 
\photoz\ ($\zphot$) since we do not  have a sufficiently large catalog 
based on \spec\ redshifts. To avoid extreme \photoz\ outliers, we remove 
galaxies that are exceedingly bright or faint for its $\zphot$ value by requiring 
a brightness-redshift constraint for
both the $r$ and $i$ band magnitudes ($m_{r,i}$),
\begin{equation}
  \mu(\zphot) - 23 < m_{r,i} <  \mu(\zphot)-16~, 
\end{equation}
where $\mu(\zphot)$
is the distance modulus for a flat $\Lambda$CDM cosmology with $\OL=0.7$ and 
$H_0=70$~(km/s)/Mpc.
This caveat has  negligible impact because the fakes are overlaid over
a wide redshift range and a wide range  of galaxy mags.

The second caveat is that the surface brightness profile is assumed to
be Gaussian (S\'ersic index = 0.5) rather than a more general
sum of S\'ersic profiles such as a bulge plus disk component.
This overly simplistic profile results in fakes 
placed preferentially near the galaxy cores with inadequate sampling 
of the disk tails. While this feature may actually help monitor
subtraction problems on galaxies, it can result in biased estimates of 
quantities that depend  on the distance to the galaxy core, 
such as measuring the fraction of SNe correctly matched to 
its host galaxy. 

The final caveat concerns masking of bad pixels. 
While the placement of fakes is independent of the masking,
the \eff\ analysis  presented here ignores fakes in which more than 10\%
of their PSF-weighted pixels are masked;
7\% of the fakes are therefore discarded.
For analyses requiring the absolute \eff, such as rates,
we can impose masking cuts on the data, or perform
additional fake studies to include the effects of masking.

\bigskip\bigskip
\subsubsection{Image Coadding and Subtraction} \label{sss:hotPants}

The program {\swarp} is used to co-add the search exposures, and to
remap the template to be aligned with the coadded search image.
For image subtraction, we use a modified version of the difference imaging program {\hotPants}.
Our version is an attempt to improve the performance, and is based on the
implementation\footnote{http://www.astro.washington.edu/users/becker/v2.0/  \\ hotpants.html}
of A. Becker, which uses the algorithm of \citet[hereafter AL98]{Alard1998},
and uses some of their original code.  The basic approach in {\hotPants}
is to  transform one image (which we call a template with pixel values $\txy$) 
so that it can be subtracted pixel by pixel from another image taken at a 
different time and under different observing conditions.
This linear transformation is described by
\begin{equation}
  t'_{x,y} = \sum^{y-y'=+r}_{y-y'=-r}\quad\sum^{x-x'=+r}_{x-x'=-r}{{k_{x(x-x')y(y-y')}
  t_{x'y'}}} \label{eqn:kernel}
\end{equation}
where $t'_{xy}$ is the convolved image which is subtracted pixel-by-pixel from the
unconvolved image.  The main computation in \hotPants\ involves the determination of
the values of the kernel of the transformation $k_{x(x-x')y(y-y')}$.  The parameter
$r$ is the size of the kernel and $x$, $y$, $x'$ and $y'$ are the pixel coordinates.
In general, one should add a constant term to Eq~\ref{eqn:kernel}, but our version 
makes a global background subtraction of the images before determining the kernel.  

The kernel is assumed to vary slowly over the image and this variation 
is described by a polynomial:
\begin{eqnarray}
  k_{x(x-x')y(y-y')}  &  =  &  k^{00}_{(x-x')(y-y')}   \nonumber \\
                              & +  & x k^{10}_{(x-x')(y-y')}  \nonumber \\
                              & +  &  yk^{01}_{(x-x')(y-y')} + ...
\label{eqn:kpoly}
\end{eqnarray}
The AL98 algorithm allows a polynomial of arbitrary order,
but since we process each CCD separately 
our \hotPants\ version includes only linear terms.

A major difference between our version and the AL98 algorithm lies in
the parameterization of the $k^{ij}_{(x-x')(y-y')}$.  AL98 parameterize
the kernel as an arbitrary number of Gaussian functions of fixed width multiplied by
polynomials whose coefficients are parameters to be fit.  
In routine use, the number of polynomial coefficients is large and comparable 
to the number of pixels  in the kernel.
Instead, we have chosen a method similar to that in \cite{Bramich2008}
in which the pixel values in the core of the kernel are fitted
without the use of a function to parameterize them.  We have, however,
retained the Gaussian function for the pixels at the edges of the kernel:  the
Gaussian form is useful for cases where a large kernel is needed to match images
with very poor seeing.  While our approach seems more transparent in terms of
understanding the fit parameters, we do not have solid evidence that our
parameterization results in better subtracted images.


\bigskip
\subsubsection{Detections and Candidates} \label{sss:detect}

We first measure the PSF to define the detection profile we 
are searching for, and then 
PSF-like objects on the subtracted image are found by {\sex}.
Selection requirements in Table~\ref{tb:obj_cuts} are applied
to reduce the number of artifacts.
An object satisfying these requirements is referred to as a ``detection."

A  ``raw \cand'' is defined when two or more detections have measured positions 
matching to within $1\arcsec$.  The two detections can be in the same band or 
different bands,  or on the same night or different nights. 
All raw \cands\ are saved, which includes moving objects such as asteroids and KBOs.  
Requiring detections on separate nights (\S\ref{ssec:pp}) is used to reject 
moving objects.

\begin{table}[h!]
\caption{  Selection Requirements on \sex\ Detections  in a Subtracted Image  }
\begin{center}  \begin{tabular}{ | l  |}    \tableline  
   (1)   ${\rm S/N} >3.5$, although the effective S/N cut from   \\
   ~~~~\sex\ is higher ($\sim 5$) as shown in Fig.~\ref{fig:fake_effSNR} \\
\hline  
  in  $35\times 35$ pixel stamp around the detected object:  \\
   (2)   fewer than 200 pixels with a flux less than $-2\sigma$ below zero,   \\
   (3)   fewer than 20 pixels with flux less than $-4\sigma$ below zero, \\
   (4)   fewer than 2 pixels with flux less than $-6\sigma$ below zero. \\
\hline 
   (5)   detection not near object in  veto catalog containing \\
   ~~~80,000 stars  with $r$-band mag $<21$.  \\
   ~~~Veto radius is mag-dependent, and  total vetoed area \\
   ~~~over all 10 fields  is  \VETOAREAsqdeg\  \degsq, or {\VETOAREApcnt}\% of the area.  \\
\hline  
  (6)   for co-added images, cosmic ray rejection based on   \\
  ~~~consistency of detected object on each exposure. \\
\hline 
  (7)  detected object profile is PSF-like based on the\\
   ~~~\sex\   
     \texttt{SPREAD\_MODEL} 
     variable \citep{Desai2012}  \\
     \hline 
   (8)   \sex\  {\tt A\_IMAGE}  $<1.5 \times {\rm PSF}$ \\
\tableline  \end{tabular} 
\end{center}   \label{tb:obj_cuts} \end{table}

\subsection{Post-processing}
\label{ssec:pp}

In addition to the single-CCD operations,
there are post-processing steps that operate on all fields and CCDs,
and continually update the \cand\  properties. 
A few percent of the events land on a CCD in two overlapping fields,
and thus single-CCD processing is not a useful concept when
constructing \cands\ from multiple \obss.

In some past surveys, as well as the start of \DESSN, the first post-processing
step was to perform a visual inspection of each detection in order
to reject subtraction artifacts that produce false detections.
In \DESSN\ we use a new machine learning based code 
to  replace human scanning;
this ``{\autoScan}'' program is described in detail in G15.
The algorithm makes use of the supervised machine learning technique Random Forest.
The training sample includes nearly 900,000 \Diff\ detections, half of which were flagged
as artifacts by human scanners and the other half are detections of fakes.
For each detection, the inputs to \autoScan\ include a $51\times 51$ pixel$^2$  
detection-centered stamp from the search, template, and subtracted images. 
The flux and \unc\ from each pixel on these three stamps contributes
$\sim 15,000$ pieces of information.
However,  rather than using the pixel-level information we found that \autoScan\ 
performs better and faster using 37  high-level features computed from the stamps. 
The three most important features are 
(1) ratio of PSF-fitted flux to aperture flux on the template image,
(2) mag-difference between the detection and the nearest catalog source, and
(3) the \texttt{SPREAD\_MODEL}  output from \sex.

For each object, the \autoScan\ program returns a score between 0 and 1,
where 0 corresponds to an obvious artifact and 1 is for a high-quality detection.
While  {\autoScan}  could have been applied before making  raw \cands, 
we have so far been conservative and apply the {\autoScan} requirement 
here in the post-processing in order to fully monitor the {\autoScan} performance.

The first post-processing step is to define ``science \cands,''
a detection on two distinct nights, each satisfying the {\tt autoScan} requirement.
Science \cands\ are the official product of  \Diff,
and as more epochs are acquired these \cands\ are repeatedly 
analyzed to select targets  (object and host galaxy)  for \spec\  \obss.
If there is a future science case requiring single-night detections,
we can recover the raw single-night \cands; 
the caveat is that during survey operations, only the 2-night science \cands\
are selected for \spec\ \obss.

The next post-processing stage is to match each science \cand\ to a host galaxy,
which is later targeted for a \spec\ redshift.
We use the ``directional-light-radius''  ($\dlr$)  method described in
\cite{Sako2014}. Currently the  galaxy profiles are approximated by a 
Gaussian (S\'ersic index = 0.5), and will eventually be updated with 
profile fits to an arbitrary Sersic index.
If there are multiple nearby galaxies within $4\times\dlr$
they are all flagged to acquire a  \spec\ redshift.

The next post-processing stage, ``forced photometry,''  computes the
PSF-fitted flux and its {\unc} for each  \obs\ since the start of the observing season,
regardless of whether there was a detection. The flux and \unc\
are computed at the same coordinates (R.A., decl.) on each subtracted image,
and the coordinates are computed as the weighted average from each detection.
This stage allows recovering small fluxes  just below detection threshold,
and fluxes consistent with zero, in order to construct complete light curves.
Ideally the \autoScan\ program
would be used to flag bad subtractions that could lead to badly
measured fluxes. However, while the \autoScan\ results exist for detections,
we do not have the infrastructure to run \autoScan\ on non-detections
in a manner analogous to the forced photometry. In addition, \autoScan\
would need additional training to accept subtractions with no significant
detection. As an alternative to \autoScan, forced photometry measurements 
are rejected from light curve fitting (below) if (1) the PSF-fitted flux and
aperture flux differ by more than $5\sigma$, or (2) within a $1\arcsec$ 
radius there are 2 or more pixels with ${\rm S/N}<-6$.

\newcommand{\Pmax}{P_{\rm max}}

The final post-processing stage is to  use the \SNANA\ program 
\PSNID\footnote{PSNID---Photometric SN Identification} \citep{PSNID}
to perform photometric classification by comparing each \cand\  light curve
to a series of photometric $griz$ light curve templates constructed on a 
redshift grid for
(1) SN~Ia,  (2) CC type II, and (3) CC type Ib/Ic. 
For each \cand-template $\chi^2$ calculation, we discard up to two epochs 
with the largest $\chi^2$ contribution (if above 10).
This outlier rejection helps to avoid bad fits from a few poorly measured
forced-photometry fluxes, particularly on bright galaxies as described in 
\S\ref{sec:monObs}.
A relative probability is computed from each $\chi^2$,
and a Bayesian probability is computed for each SN type;
the largest probability ($\Pmax$) determines the type and redshift.
If $\Pmax < 0.5$, or the best fit $\chi^2$ is poor, 
the candidate type is flagged as unknown.
The  probability for each type and the estimate of peak magnitude
contribute to the \spec\ target selection process
(\S\ref{sss:liveSpec}).

\subsection{Spectroscopic Target Selection}
\label{subsec:spec}

While \spec\ target selection is outside the scope of \Diff,
here we give a brief description to give a more complete picture
of the \DESSN\ program. The two components of \spec\ targets,
host redshifts and live transients, are described below.

\subsubsection{Host Galaxy Redshifts}
\label{sss:zSpec}

The large numbers and faint magnitudes of SNe discovered in \DESSN\  overwhelm the
available resources for  \specy\ classifying each candidate.
However, we can efficiently use multi-fiber \spec\ 
resources to measure an accurate host-galaxy redshift for the majority of
our SN \cands. 
Using the Anglo-Australian Telescope (AAT), the OzDES program \citep{OzDES}
is a 100-night spectroscopic survey 
with the 400 fiber Two Degree Field (2dF) instrument 
feeding the dual-beam AAOmega spectrograph.  
The overlap between the field of view of DECam and 2dF is nearly complete.  
With repeat visits to the same source,  spectra are coadded to enable
redshift measurements for much fainter galaxies than would naively 
be expected from a 4~m class telescope;  
redshifts are obtained for about half of the  24th mag galaxies ($r$ band).
In addition to targeting host galaxies for SN \cands,  OzDES also 
targets a variety of DES sources such as
AGN to derive reverberation mapped black-hole masses,
galaxies for DES photo-$z$ calibration, 
white dwarfs for calibration,
and live transients for \spec\ typing.

\subsubsection{Spectroscopic Identification of Live Targets}
\label{sss:liveSpec}

The \spec\ selection for live transients is primarily focused
on SN~Ia. The selection is based on a visual examination of 
light curves along with \PSNID\ probabilities.
The phase estimate is used to give higher priority 
to \cands\ near peak brightness.
Highest priority is given to  \cands\ with peak $r $ band magnitude 
$r_{\rm  peak} < 20.5$~mag (mag-limited)
and to \cands\ with a photometric redshift below 0.2 (volume limited).
These two samples have large overlap, and are expected to be
very nearly complete.
Lower priority is given to \cands\ over the full redshift range
where we expect to acquire a \spec\ typing for $\sim 10\%$
of the SN~Ia sample. 
Starting in Y2, transient activity in multiple seasons is used
to reject AGN-like \cands.

Telescopes used to \specy\ confirm transients discovered by \Diff\ include
the 3.9-m Anglo-Australian Telescope (AAT) at Siding Springs Observatory in Australia,
the 8.2-m Very Large Telescope (VLT) on Cerro Paranal in Chile,
the 9.2-m South African Large Telescope (SALT) near Sutherland in South Africa,
the 10.4-m Gran Telescopio Canarias (GTC) in La Palma,
the Keck 10-m on Mauna Kea in Hawaii,
the 6.5-m Magellan Telescope at Las Campanas Observatory in Chile,
the 6.5-m MMT on Mount Hopkins in Arizona,
the 3-m Shane telescope  at Lick Observatory in California,
the 4.1-m Southern Astrophysical Research (SOAR) telescope at Cerro Pachon in Chile,
the 8.1-m Gemini-South telescope at Cerro Pachon in Chile,
and
the 9.2-m Hobby-Eberly Telescope at the McDonald Observatory in Texas.

\subsection{ Statistics Summary }
\label{ssec:diff_stats}

A summary of the first-season (Y1)
statistics for single-epoch detections is shown in 
Table~\ref{tb:obj_stats}.  The average number of objects per field
found by \sex\ increases with the passband central  wavelength.
In the shallow fields there are
$\sim 100,000$ per field in the $g$ band,
increasing to $\sim 170,000$ in the $z$ band.
In the deep fields there are 130,000  in the $g$ band, increasing to
270,000 in the $z$ band.  A visual scanning assessment shows
that more than 90\% of these detections are subtraction artifacts.
Following the \sex\ detections on the subtracted image,  
there is a significant reduction from the selection cuts and \autoScan. 
The selection cuts reduce the number of detections by a factor
of 3-4 in the $g$ band, and a factor of $\sim 2$ in the $z$ band.
The automated scanning provides a further reduction of a factor
of $\sim 4$ in the $g$ band, increasing to an order of magnitude in the $z$ band.
After all selection requirements and automated scanning,
the average number of objects per field in Y1 is  $\sim 10^4$
in both the deep and shallow fields, 
and the artifact fraction is $\sim 25\%$
as determined from a visual scanning assessment.

To determine the average number of detections per square degree
for a single-epoch visit ($\ndetect$), the number of Y1 detections 
(\autoScan\ row in Table~\ref{tb:obj_stats})
is divided by 2.7~\degsq\ and $\Nvisit$ from Table~\ref{tb:fields}:
$\ndetect \approx \NDETECTg$ in the $g$ band and $\approx \NDETECTz$ in the $z$ band.

\begin{table} 
\caption{
  Number of Non-fake Single-Epoch  Detections  in the Y1 Season
  per 3~\degsq~Field (thousands)
  } 
\begin{center}
\begin{tabular}{ | l | l | cccc | }
\tableline
               &          &  \multicolumn{4}{c |}{ Number of Detections }  \\
              &          &  \multicolumn{4}{c |}{ $(\times 10^{3})$ }  \\
              &  Detection       & \multicolumn{4}{c |}{ }        \\
 Fields     &  Stage             &   $g$  &  $r$  &  $i$  &  $z$       \\
   \hline\hline   
   Deep  &    \sex\                               & 133 & 166 & 277 & 270      \\
               &   + selection cuts\tablenotemark{a}
                                                                          &  32 &  81 & 172 & 167    \\
              &   + {\autoScan}\tablenotemark{b}     &   8 &   8 &   9 &  12      \\
             &   \autoScan/cuts ratio        & 0.25 & 0.10 & 0.06 & 0.07  \\
 \hline 
    Shallow    &  \sex\                               & 98 & 103 & 126 & 173       \\  
                    &  + selection cuts\tablenotemark{a}
                                                                                  &  29 &  26 &  55 &  92   \\ 
                    &  + {\autoScan}\tablenotemark{b}        &   8 &   7 &   9 &  10   \\ 
                    &  \autoScan/cuts ratio        & 0.28 & 0.27 & 0.18 & 0.11   \\ 
\tableline  
\end{tabular}
\end{center}
 \tablenotetext{1}{Includes selection cuts in Table~\ref{tb:obj_cuts}.}
 \tablenotetext{2}{Includes cuts and automated scanning requirement (G15).}
  \label{tb:obj_stats}
\end{table}

The total number of raw \cands\ in Y1, which requires two \sex\ detections passing 
the selection cuts in Table~\ref{tb:obj_cuts}, is \NRAWCANDTOT.
Requiring two detections on different nights reduces this slightly to \NTMPCANDTOT. 
Requiring the two separate-night detections to satisfy the automated scanning reduces 
the number of candidates to \NSCICANDTOT,  or a factor of {\autoScanRej} reduction.
Table~\ref{tb:cand_stats} shows the average number of \cands\
per deep field and per shallow field.

Following \sex\ detections, the selection cuts and \autoScan\ have
a dramatic effect on reducing the number of detections and \cands.
This is because the vast majority of the \sex\ detections are 
false positives, or artifacts of the image subtraction. These
artifacts come from a variety of sources, including 
bright stars and galaxies,
defective pixels,
edges of masked regions,
CCD edges, and
cosmic rays.
Some of these artifacts are illustrated in Fig. 1 of G15.
The large rejection by \autoScan\  costs only a 1.0\% 
loss of fake SNe~Ia \cands, mainly for fakes with low S/N at peak brightness.
We are  therefore confident that \autoScan\ is highly efficient for real 
astrophysical transients.

\begin{figure*}   
\epsscale{1.1} 
\plotone{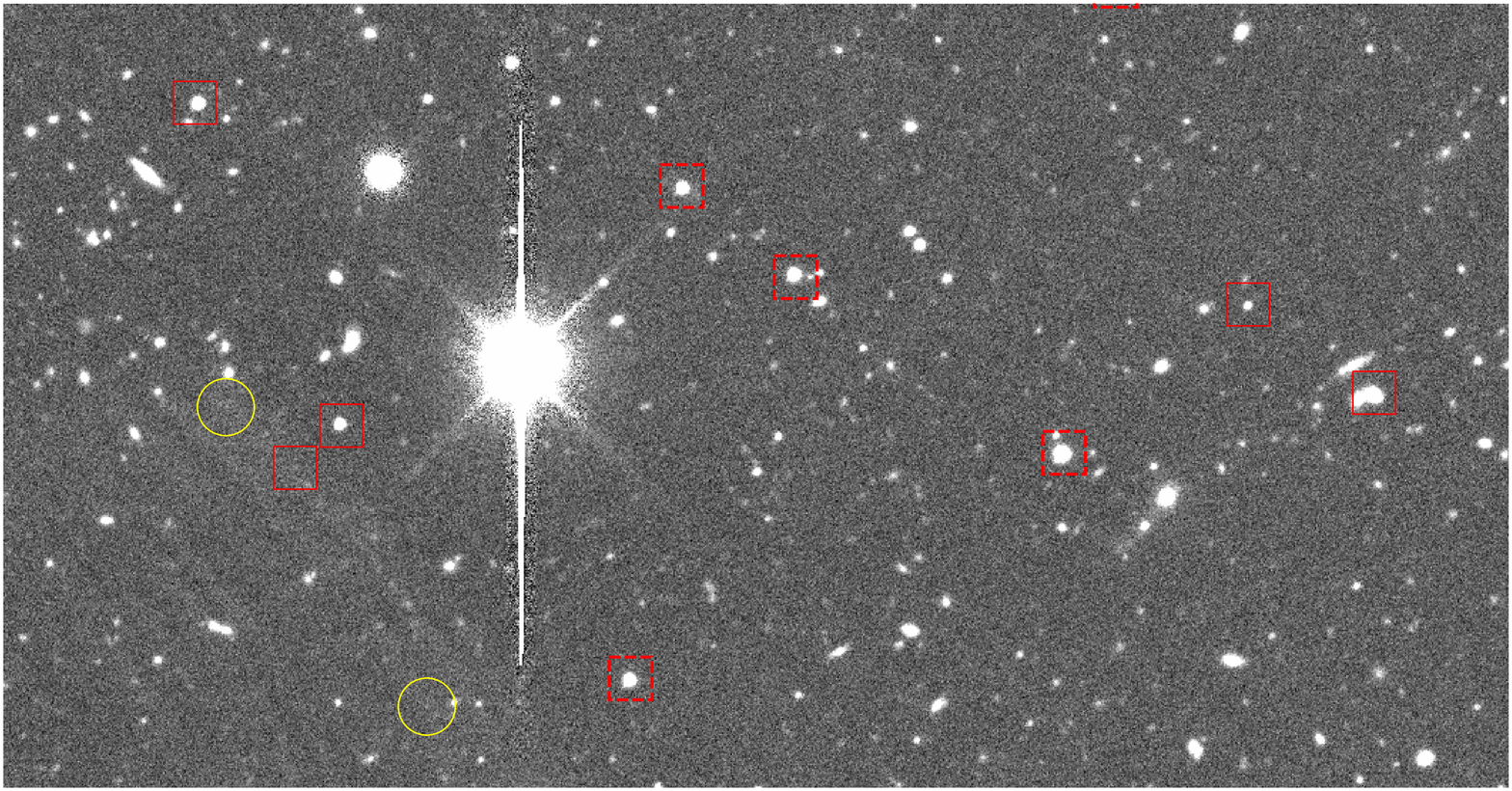}
\plotone{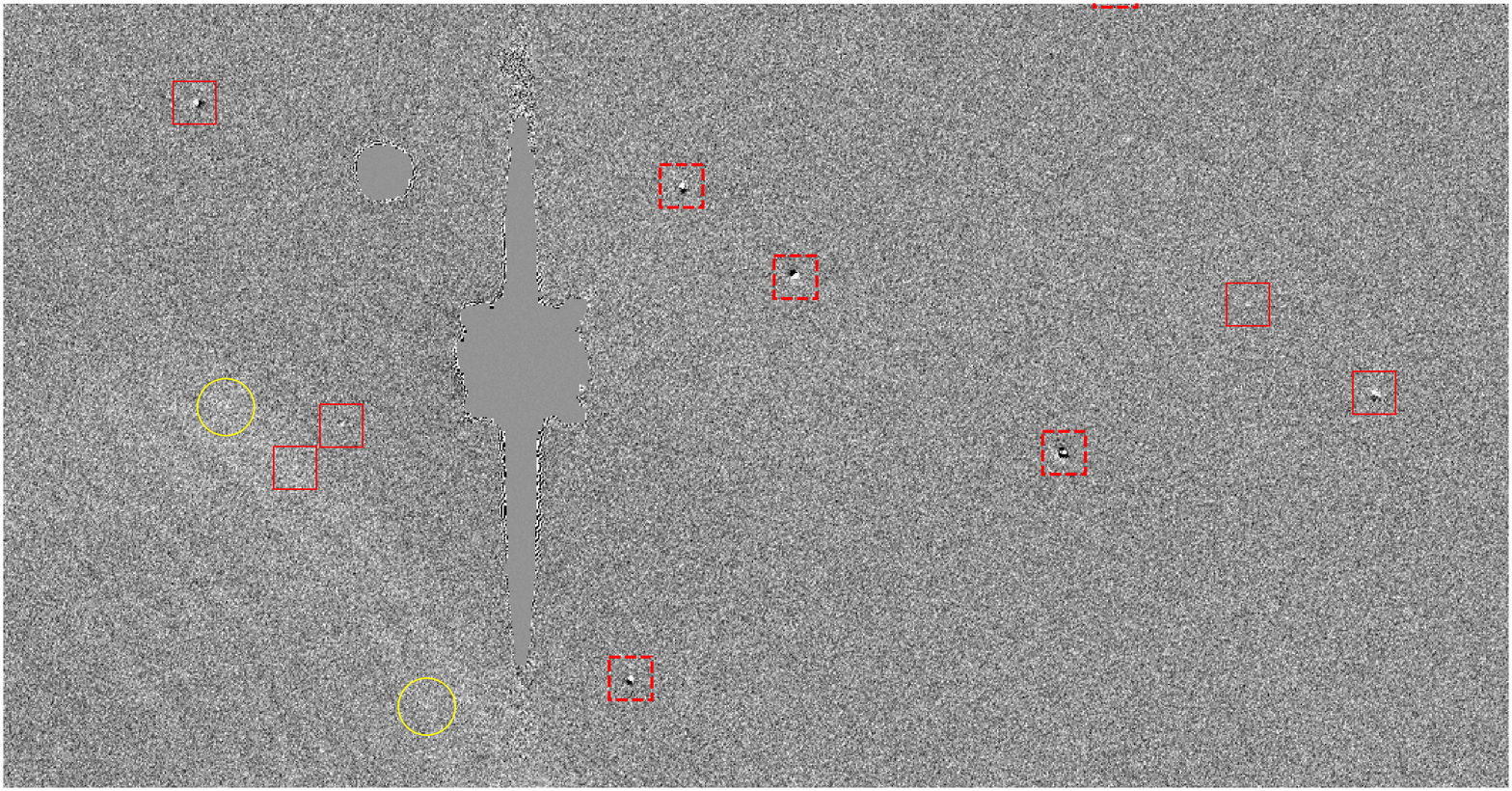}
\caption { 
	From \Diff,
	a co-added search image (top) and subtracted image (bottom) from a typical night
	(2013 October 13) in deep field C3 for  $i$ band.
	The image size is roughly $2.7\arcmin \times 4.6\arcmin$, 
	or about 1/13 the area viewed by a single CCD.
	Non-fake \sex\ detections on the subtracted image (bottom) 
	are highlighted in both images:
	dashed red boxes for objects failing the selection cuts in Table~\ref{tb:obj_cuts},
	solid red boxes for objects passing these cuts and failing \autoScan, and
	yellow circles for objects passing cuts and \autoScan, which are used
	to make science \cands.
	To set the scale, the brightest masked star has mag $m=11.4$;
	the other masked star has mag $m=15.0$.	
	To see detections in more detail, Fig.~1 in G15 shows a collection of 
	$51\times 51$~pixel$^2$ stamps for search+template+subtracted images, 
	each centered on a detection.
	}
  \label{fig:ds9}
\end{figure*}

Subtraction artifacts are illustrated in Fig.~\ref{fig:ds9} for a deep field 
image processed by  \Diff.
\sex\ detections failing selection cuts (dashed red boxes) are the
most clearly evident upon visual inspection, while those failing
\autoScan\ (solid red boxes) are more subtle. In this example,
most of the artifacts are around a few bright objects even though most 
of the bright sources are cleanly subtracted. On average, artifacts
are $\sim 1$~mag fainter than real transients.
To get an estimate of the artifact rate for bright sources, 
$\sim 3$\% of bright fakes (mag$<20$) fail the detection 
and {\autoScan} requirements.
The origin of these artifacts is not understood.

\begin{table}  
\caption{
  Average Number of Non-fake Y1-\cands\ per 3~\degsq~Field.
  } 
\begin{center}
\begin{tabular}{ | l  |  cc |}
\tableline  
      Candidate   & \multicolumn{2}{c | }{ $\Ncand$  per Field }  \\
       Selection    &      DEEP   &  SHALLOW          \\
   \hline\hline 
  2 detections (raw cand)      &  18830 & 10410 \\  
  2 nights (without \autoScan) &  17460 & 8230 \\  
  2 nights + \autoScan\ (science cand) &  \NcandPerDeep\  & \NcandPerShallow\   \\  
\tableline  
\end{tabular}
\end{center}
  \label{tb:cand_stats}
\end{table}

\subsection{Classification Summary}
\label{subsec:psnid_summary}

\newcommand{\MJDcand}{{\rm MJD}_{\rm cand}}
\newcommand{\MJDref}{{\rm MJD}_{\rm ref}}

Here we show the breakdown of \PSNID\ classifications for science \cands\
(\S\ref{ssec:pp}).
To avoid the noisiest light curves we consider the subset in which three bands 
each have an \obs\ with S/N$>5$; this subset is roughly half of all \cands.
Applying \PSNID\ to the entire light curves for the full
Y1 sample results in nearly equal classification 
fractions ($\sim 1/3$) for SN~Ia, SN~CC (mostly Type II) and unknown.

\begin{figure}   
\plotone{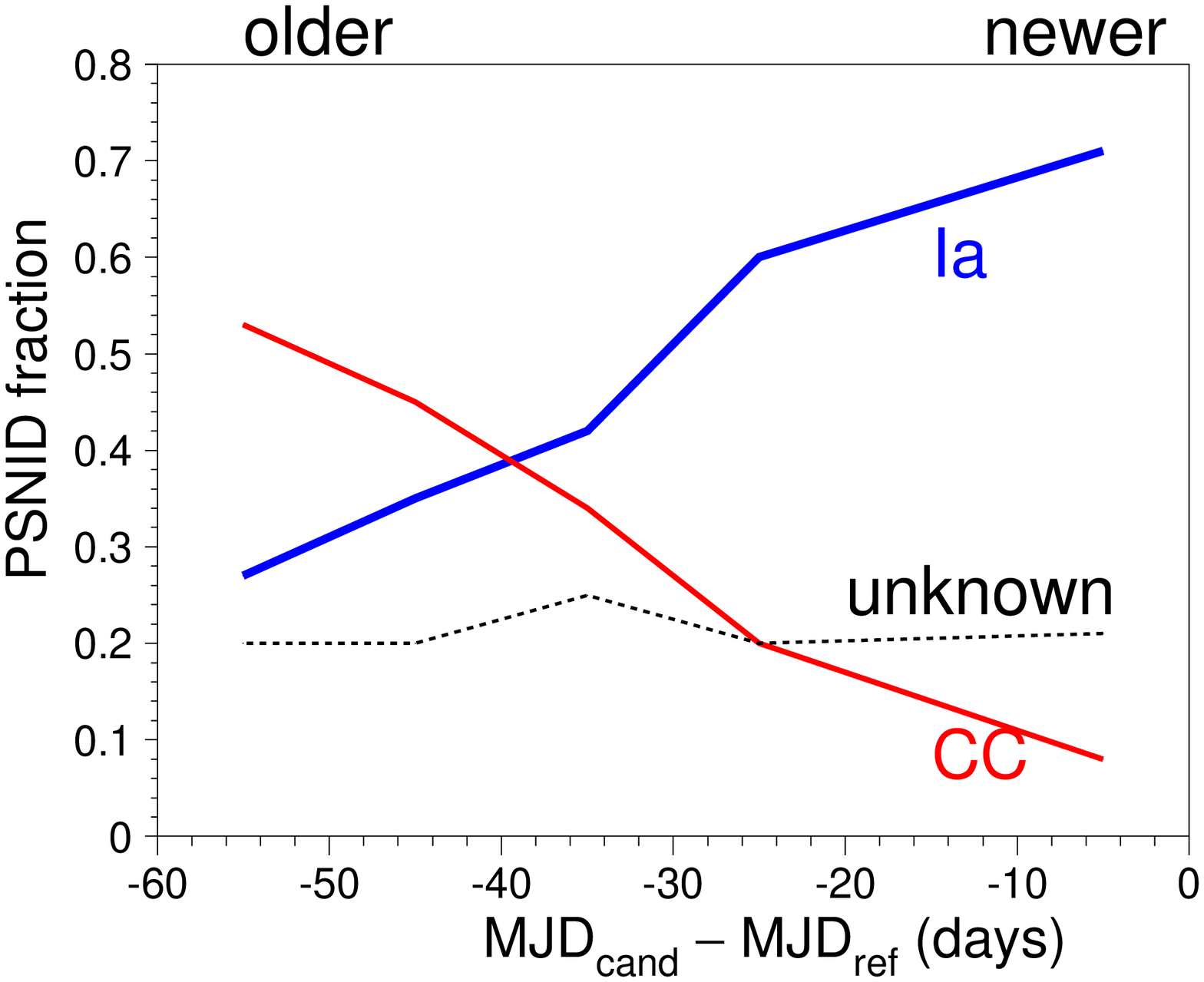}
\caption { 
     \PSNID\ classification fraction for SN~Ia and SN~CC
      vs. time that the light curve has been observed,
     for \cands\ in which three bands have an \obs\ with S/N$>5$.
     See \S\ref{subsec:psnid_summary} 
      for explanation of $\MJDcand$ and $\MJDref$.
     	Zero on the horizontal axis corresponds to newer
	\cands\   used in the \PSNID\ fits;   
       $-50$ days corresponds to older \cands\ whose
       second detection occurred 50 days earlier
       and thus have longer light curve coverage in the \PSNID\ fits.
	``Unknown" corresponds to light curves for which \PSNID\ cannot
	determine a SN type.
}
\label{fig:psnid_frac} \end{figure}

While a full Y1 analysis is relevant  after the survey,
during survey operations \PSNID\ is run on newly
discovered light curves that have only a few epochs.
To illustrate the real-time \PSNID\ performance,  Fig.~\ref{fig:psnid_frac} 
shows the classification fractions as a function of time the light curve 
has been observed. 
$\MJDcand$ is the time when the second epoch is detected,
or when the object became a science \cand.  
$\MJDref$ represents the current MJD, which we take to be 56,600 in this example.
The fits include \obss\ between $\MJDcand-20$ and $\MJDref$.
When only the early part of the light curve is available for fitting
($-5$~days in Fig.~\ref{fig:psnid_frac}),
about 70\% of the \cands\ are classified as SN~Ia, fewer than 10\% as SN~CC,
and the rest are unknown.  When fitting 2~months of the light curve,
more than half of the  classifications are SN~CC.

\bigskip\bigskip
\subsection{ Data Reprocessing }
\label{subsec:reproc}

During Y1, the monitoring of fakes showed a significant \ineff\ that was 
traced to severe astrometry problems as described in \S\ref{sss:astrom}.
This problem was fixed after Y1, 
and before the start of Y2 operations all of Y1 was reprocessed in order to recover
hundreds of host-galaxy \spec\ targets that had been missed during Y1.
During Y2, the monitoring of fakes showed good \Diff\  performance
in the shallow fields, but there were still significant flux-outliers
in the deep fields. This problem was eventually traced to 
the program which determines the PSF used for 
calculating PSF-fitted fluxes, and it was fixed after Y2.

Both Y1 and Y2 have been fully reprocessed in all ten SN fields, with all \Diff\ fixes.
Results presented in this paper are based on the Y1 season,
using templates constructed from Y2 images.
The reprocessed results are used
to discover transients missed during the survey, to 
update the photometric classification with \PSNID,
and to update the host-galaxy target list for measuring
\spec\ redshifts.
While transients discovered in the reprocessing have become too
faint to target for \spec\ \obss,
this is not a serious issue because we target only a small fraction 
of the transients anyway.

Another subtle change in the reprocessing campaign was to 
fully analyze each exposure in the deep field sequences 
(in addition to the coadd) to improve the KBO search.
In particular, this reprocessing led to the discovery of
one of the two Neptune Trojans in \citet{KBO2015},
as well as improved orbital fits for both objects.
We are currently upgrading \Diff\ to overlay fake KBOs onto the images; 
these fake KBOs will allow measuring the search \eff,
and they will be used to develop improved KBO-finding algorithms.

We do not expect more \Diff\ improvements during the remainder of DES, 
unless our monitoring uncovers new  problems or we improve the 
subtraction problem on bright galaxies as described in \S\ref{subsec:SB}.
Even without software changes, we may reprocess the data
in the future using better templates and  lower detection 
thresholds in order to improve the depth of the search.

\subsection{ \Diff\ Processing Time }
\label{subsec:cpu}
Using the IBM iDataPlex Carver computational system at 
NERSC\footnote{National Energy Research Scientific Computing Center}, 
we give the processing time for the \Diff\ steps in the middle 
panel of Fig.~\ref{fig:diffim_overview}.
For a shallow field with a single exposure, the processing time for a single CCD is
$\sim 10$~minutes, half of which is spent on the \hotPants\ program. 
In the deep fields we  perform the \hotPants\ subtraction for each exposure
as well as the coadded image, and thus the processing time scales roughly
with the number of exposures. For a deep-field sequence with 11 $z$ band
exposures, the processing time for a single CCD is $\sim 90$~minutes.

The post-processing steps (right panel in Fig.~\ref{fig:diffim_overview}) 
run serially, and the processing time depends on how long the survey has been running. 
Near the start of a survey season the post-processing takes a few minutes, 
but near the end of the season it takes several hours.

\section{{\Diff} Monitoring-I: Single Epochs}
\label{sec:monObs}

Here we describe monitoring of the single-epoch detection \eff\  and data quality,
using both the MAG20 fakes and the SN fakes processed by \Diff.

\subsection{ Data Quality Assessment}
\label{subsec:dataq}

\begin{figure}[h!]   \centering
\epsscale{1.1} 
\plotone{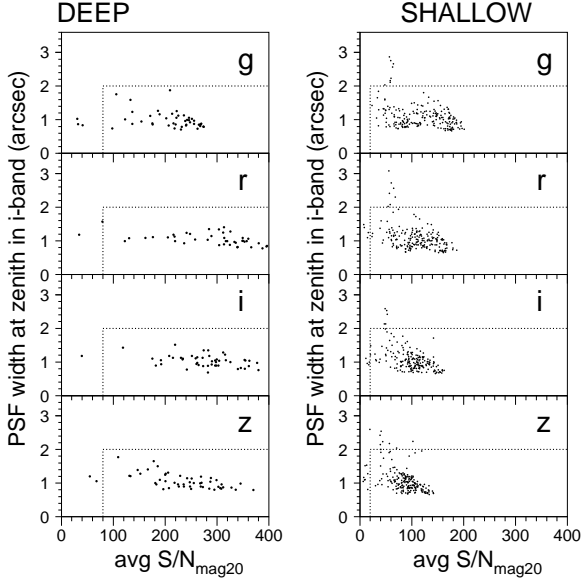}
\caption { 
  	For each set of exposures in Y1, the CCD-averaged PSF width 
	($i$-band at zenith, FWHM, arcsec) 
  	is plotted against the CCD-averaged S/N from the MAG20 fakes.
      Exposure sequences with points outside the dashed box are retaken,
      typically the following night.
      Left panels are for the deep-fields  and the right panels are for the shallow fields.  
	}
  \label{fig:SNR_MAG20}
\end{figure}

The measured S/N
from the MAG20 fakes is part of the data quality evaluation (See Fig.~\ref{fig:SNR_MAG20}).
We define $\SNRvingt$ to be the average S/N among all of the ($4\times 60=240$) 
MAG20 fakes overlaid on each exposure, where each S/N is the ratio of the 
PSF-fitted flux to its \unc.
If $\SNRvingt < 20$ in the shallow fields, or $< 80$ in the deep fields, 
the exposures are flagged to be retaken.
In addition, an exposure is retaken if the $i$ band PSF width (FWHM) at zenith 
is $>2\arcsec$; this 
seeing value is computed by correcting the measured PSF for airmass and wavelength. 
These criteria for retaking an exposure are a compromise between 
data quality in the SN fields and lost observing in the wide-area survey.
The largest $\SNRvingt$ values are from high-quality data triggered 
because there were no \obss\ within the past 7 days.
The lower $\SNRvingt$ values are typically from  data triggered by seeing 
$>1.1\arcsec$ and from \obss\ at larger airmass.

\begin{figure}[h!]   \centering
\epsscale{1.1} 
\plotone{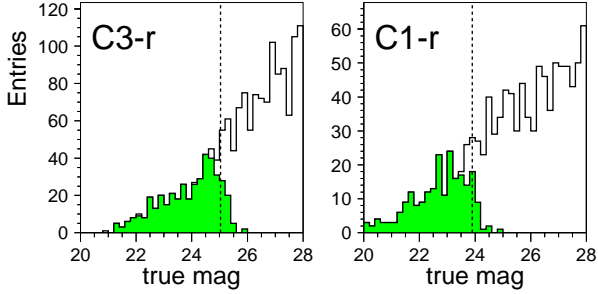}
\caption { 
      True mag distribution for fakes from a single epoch.
      Left panel is for deep field C3-$r$; right panel is for shallow field C1-$r$.
      Shaded overlay is for fakes satisfying \Diff\ detection requirements.
      The dashed vertical line shows $\mhalf$ as defined in the text.
	}
  \label{fig:effTrueMag}
\end{figure}

While $\SNRvingt$ and the PSF are used to determine
if an exposure sequence needs to be retaken, the SN~Ia fakes are used
to determine complementary information about the data quality.
For a given epoch, the fakes are used to determine the magnitude depth, 
$\mhalf$, defined as the mag where the \Diff\  detection \eff\ has fallen to 50\%.  
Figure~\ref{fig:effTrueMag} illustrates the determination of $\mhalf$.
The $\mhalf$ distribution is shown in Fig.~\ref{fig:mageff50}
for each band, and for deep and shallow fields. The variation in
$\mhalf$ is from the variation in observing conditions.

\subsection{Detection Efficiency vs. S/N}
\label{subsec:eff_vs_snr}

The detection \eff\ as a function of S/N ($\effSNR$) is a crucial 
input to the MC simulation (\S\ref{sec:monCand}) and also 
provides another monitoring metric.
We do not attempt a first-principles calculation of $\effSNR$,
primarily because of the complicated behavior of \sex\ 
that largely defines the detection threshold. 
Therefore $\effSNR$ is empirically measured from the fakes as illustrated in 
Fig~\ref{fig:fake_effSNR} for the $i$ band.
The effective S/N threshold, defined for $\effSNR = 0.5$, is about 5 in each band
and is the same in both the deep and shallow fields.
Each sub-panel shows the nominal $\effSNR$ curve 
computed from all of the fake data, along with a systematic test based
on splitting the data into two equal-size samples. 
The probability of detecting a transient depends on the ZP, PSF, and 
sky-noise through their effect on the S/N,  and we expect the detection \eff\ 
to depend primarily on S/N.
Fig~\ref{fig:fake_effSNR} shows that there is no unexpected dependence,
which is important because not all of the selection criteria are based on S/N.

\subsection{Anomalous Subtractions on Bright Galaxies}
\label{subsec:SB}

The final issue is the reliability of  forced-photometry flux measurements that 
are used to classify light curves, both visually and with fitting programs.
The average fake fluxes are recovered to within  few percent of their 
true values, which is adequate precision since it is smaller than the
model errors used in light curve fitting.
We have also checked the reliability of  the flux \uncs,
and found that these \uncs\ are underestimated  in proportion
to the local galaxy surface brightness (SB) under the SN location;
we refer to this effect as the ``\SBa.''
The  excess flux scatter can cause problems with monitoring and
light curve fitting, and thus we have modeled this effect 
in both simulations and  fitting programs (\S\ref{sec:monCand}).

To define the SB, we first sum the template flux at the
\cand\ location, using an aperture with $1.3\arcsec$ radius,
which contains most of the flux for a typical PSF.
The SB flux is defined as the average flux per square arcsecond,
and the SB-mag ($\mSB$) is the corresponding magnitude per
square arcsecond. 
For fakes we characterize the quality of the \uncs\ using the 
rms of $\DF / \sigF$ ($\rmsD$), 
where $\DF$ is the difference between the measured  (forced photometry) 
flux and the true flux of the fake, and $\sigF$ is the \unc\ 
on the forced-photometry measurement.
Ideally $\rmsD=1$ in all cases, but we find that $\rmsD$ increases
with SB as shown in Fig.~\ref{fig:pullSB_deep} for the deep fields
and in Fig.~\ref{fig:pullSB_shallow} for the shallow fields.
For low SB ($\mSB > 24$),  $\rmsD$ is very close to unity
as expected. For the brightest galaxies where $\mSB \approx 20$,
$\rmsD \approx 5$ in the deep fields and $\sim 3$ in the shallow fields.

Figures~\ref{fig:pullSB_deep} and \ref{fig:pullSB_shallow}  also show 
rms vs. $\mSB$ separately for dim fakes with $m>26$ (red curve) 
and for brighter fakes with $m<24$ (blue curve).
The consistency shows that this effect depends mainly on the brightness
of the galaxy and not the transient source.

\begin{figure}[h!]   \centering
\epsscale{1.1} 
\plotone{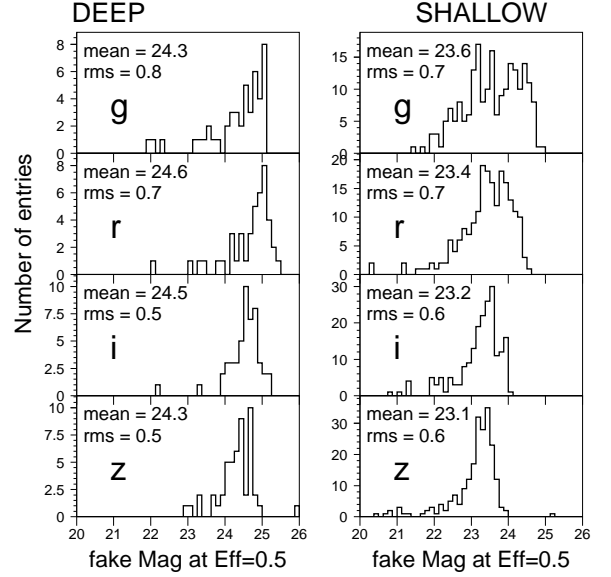}
\caption { 
      Distribution of $\mhalf$ in each passband, determined with fakes.
      Each entry is from one epoch. Left panels are for the
      deep fields; right for the shallow fields.
	}
  \label{fig:mageff50}
\end{figure}

\begin{figure}  
\centering
\plotone{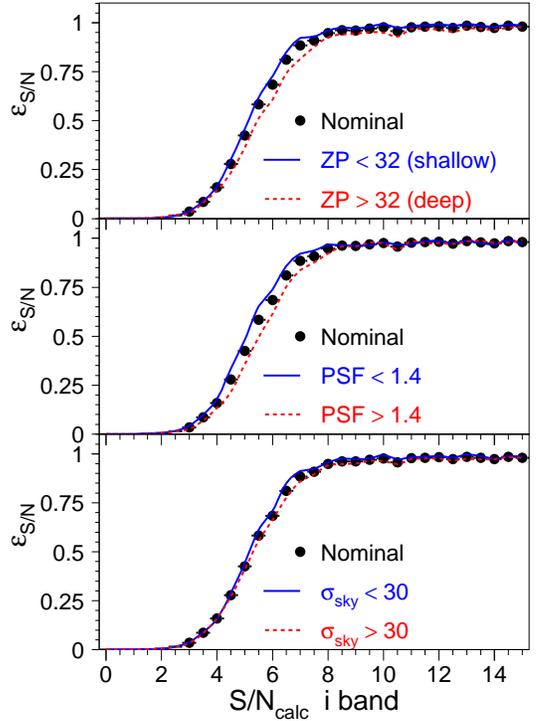}
  \caption{
   Single-epoch detection \eff\ ($\effSNR$) vs.  S/N, as measured with fakes.
   The solid-filled circles are computed from all the data, and is the same in each panel.
   The solid and dashed curves correspond to splitting the sample into roughly 
   two equal bins for zero point (top), PSF (middle) and sky noise (bottom).
  }
  \label{fig:fake_effSNR}
\end{figure}

\begin{figure}  
\centering
\plotone{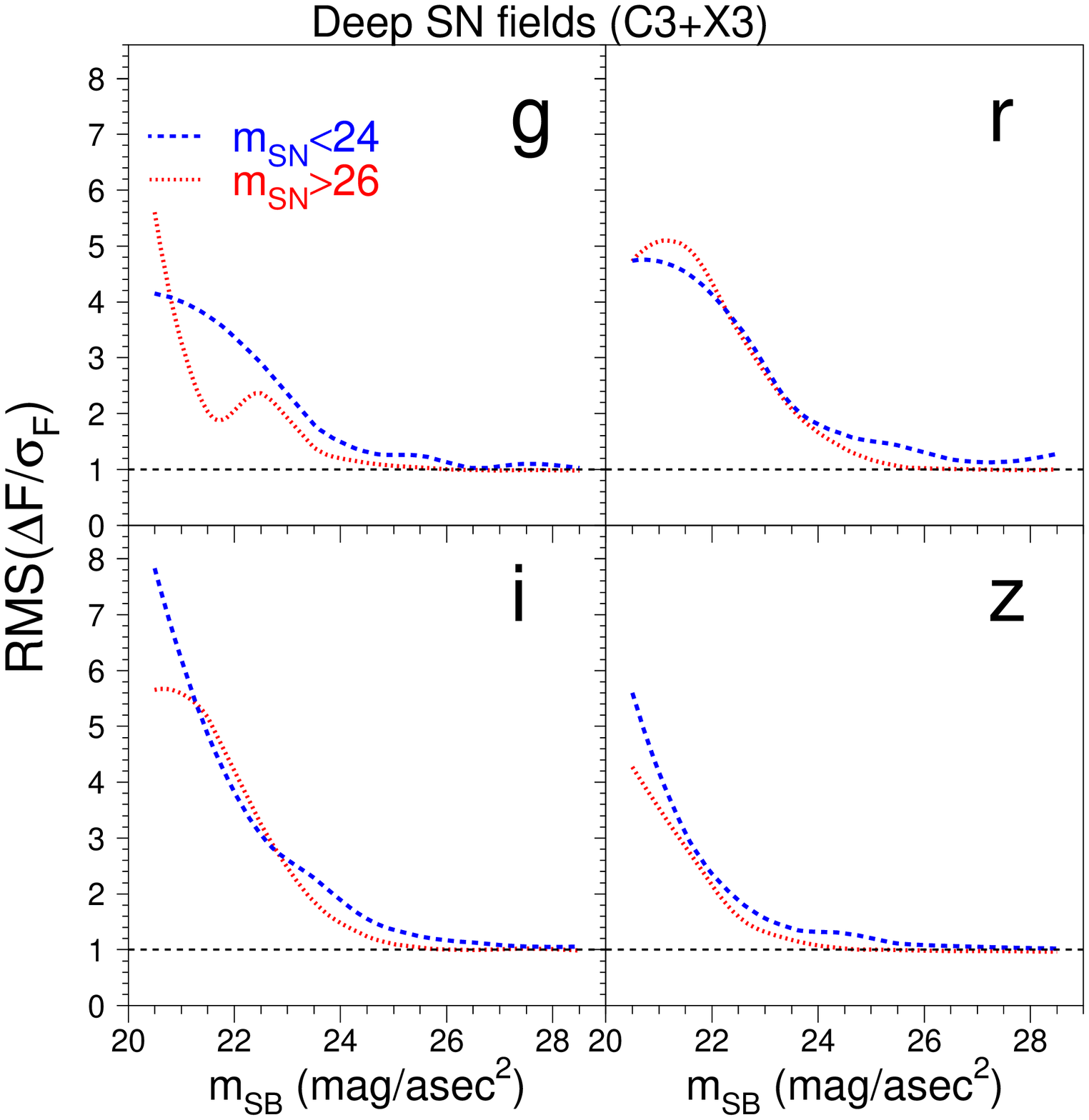}
  \caption{
    For the 2 deep fields in each pass band, 
    RMS of $\DF/\sigF$ as a function of the galaxy surface-brightness mag ($\mSB$)
    for fakes as defined in the text.  $\DF$ is the difference between the true and
    measured fake flux, and $\sigF$ is the \unc.
    The horizontal dashed line through 1 shows the expected value
    if \Diff\ correctly determines the flux \uncs. 
    The dotted-red curve is for very faint fakes with SN mag $m>26$; 
    the dashed-blue curve is for fakes with SN mag $m<24$.
    The model for this effect (\S\ref{sec:monCand}) depends on $\mSB$
    and not on the transient mag.
  }
  \label{fig:pullSB_deep}
\end{figure}

\begin{figure}  
\centering
\plotone{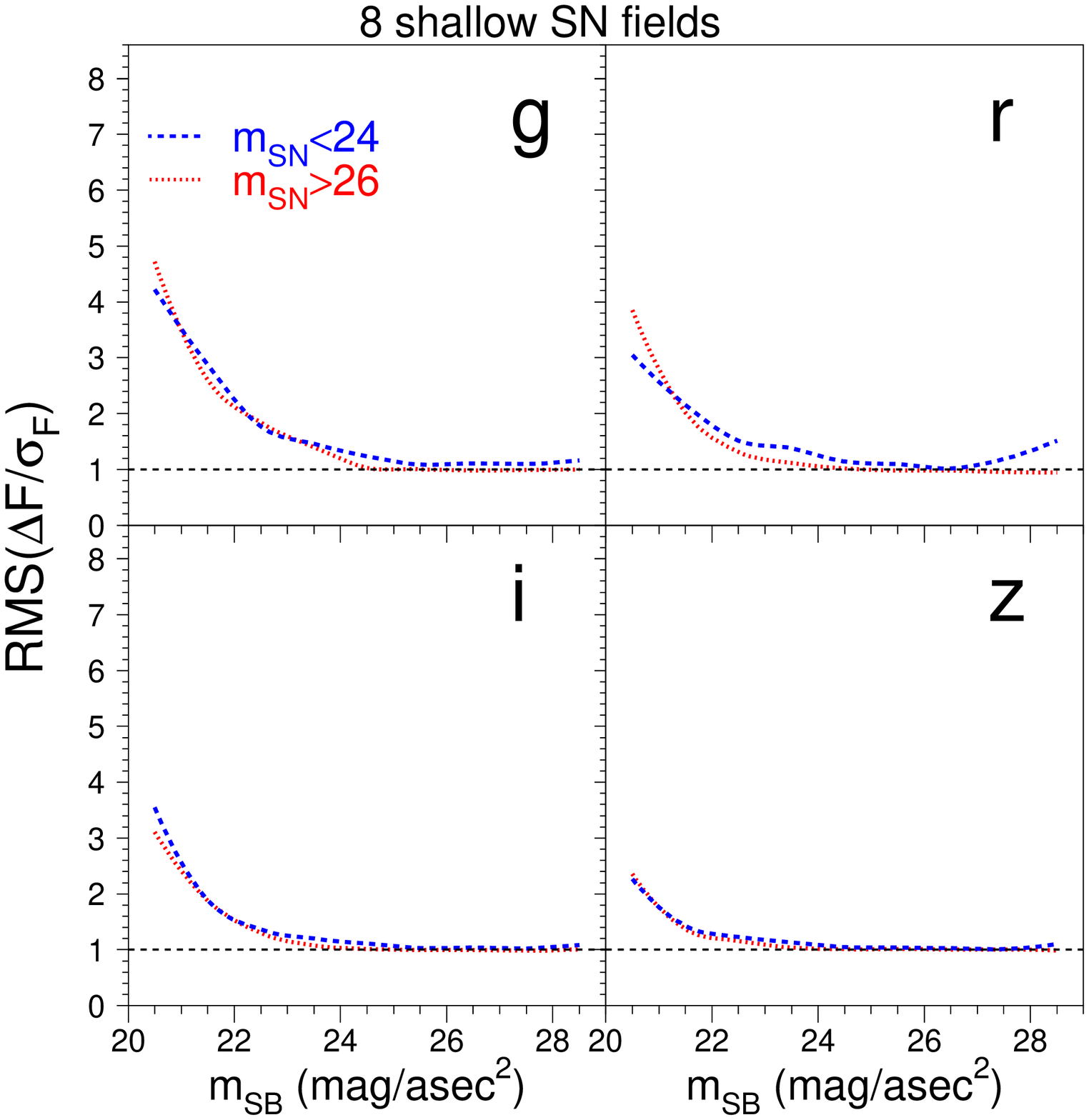}
  \caption{
	Same as Fig~\ref{fig:pullSB_deep}, but for the 8 shallow fields.
	}
  \label{fig:pullSB_shallow}
\end{figure}

\section{ {\Diff} Monitoring-II:  Science Candidates}
\label{sec:monCand}

While monitoring the single-epoch detection \eff\ and data quality
are important on a nightly basis (\S\ref{sec:monObs}), the science prospects
ultimately depend on the \Diff\ \cand\ \eff\ and our ability to 
select \spec\ targets based on a small number of epochs.
Here we describe the \Diff\ monitoring  of science {\cands}
using SN fakes combined with MC simulations. 
The basic idea is to use the MC simulation to predict the SNe~Ia \eff\ versus redshift, 
and compare with the true \eff\ measured from the fakes. 

There are two different \effs\ to monitor as a function of redshift.
The first  \eff\  is the fraction of fakes that become a science \cand\ ($\effCand$). 
As long as $\effCand$ is optimal, then even if  \Diff\ measures fluxes with
many catastrophic outliers an improved offline photometry analysis can make
all of the discovered light curves useful for science analysis.
However, if there are too many flux outliers
then real-time photometric classification becomes
more difficult, which complicates the selection of \spec\ targets.

It is therefore important to monitor a second \Diff\ \eff,
the fraction of fakes passing the photometric analysis ($\effPSNID$)
used for \spec\ targeting,
which is based on the  \PSNID\ program \citep{PSNID}.
The key component of the \PSNID\ analysis  (Table~\ref{tb:PSNID_CUTS})
is a requirement on the fit probability computed from the
template-fit $\chi^2$,  and therefore even a few 
measured fluxes that are highly discrepant from their true values
can cause \PSNID\ to reject the light curve. 	
Up to two highly discrepant fluxes (w.r.t. the fit) are rejected,
allowing for a small level of subtraction problems.
In summary, simply discovering an event is not adequate unless the 
flux measurements  are of sufficient quality to 
perform light-curve template fitting without suffering significant \ineff.

\begin{table}[h!]
\caption{PSNID Analysis Requirements for $\effPSNID$ Study }  
\begin{center}  \begin{tabular}{ |  l   |  l  |  }    \tableline  

   Category       &  Requirement  \\
      \hline\hline 
  \hline
    Sampling  & 5 or more \obss.  \\
                     & 3 bands with at least one S/N$>5$ \obs. \\
                     & An \obs\ with $\Tobs < -2$ days.\tablenotemark{a} \\
                     & An \obs\ with $\Tobs > +5$ days.  \\
  \hline 
    Fit-$\chi^2$  & Fit prob $\Pfit > 0.1$\tablenotemark{b} \\
                       & Reject up to two $3.16\sigma$ data-fit outliers ($\Delta\chi^2>10$). \\   
   \hline
   Typing   & Best fit template (among Ia, II, Ib, Ic) is Type Ia. \\
\tableline  \end{tabular} 
 \tablenotetext{1}{$\Tobs$ is the observer-frame time since the epoch of peak brightness.}
 \tablenotetext{2}{
   $\Pfit$ is calculated from $\chi^2$/dof.
   Because of the large \PSNID\ model errors, the true chance of
   finding $\Pfit < 10\%$ for SNe~Ia is $\sim 1$\%.}
\end{center}   \label{tb:PSNID_CUTS} \end{table}

Details of the MC simulation  are given  in Appendix~\ref{app:mc},
and here we give a brief overview.
The MC simulation uses the observed cadence,
and the simulated flux and noise are computed from the
observing conditions at each epoch:  ZP, PSF, sky noise, CCD gain.

While the cadence information is trivially obtained from survey \obss,
the MC simulation also needs two inputs based on the fakes processed by  \Diff.
First, we  use the  \eff\ vs. S/N  ($\effSNR$)  measured in each passband, 
and illustrated in Fig.~\ref{fig:fake_effSNR} for the $i$ band.
Since there is good agreement between the deep and shallow fields,
we use the same $\effSNR$  function in all fields.

The second input from the fakes is a model for the \SBa,
the anomalous flux \unc\ that increases with the local surface brightness. 
The galaxy Sersic profile in the simulation is used to analytically compute $\mSB$, 
and the $\rmsD$-versus-$\mSB$ curves in 
Figures~\ref{fig:pullSB_deep} and \ref{fig:pullSB_shallow}
are used to scale the sky noise as a function of passband,
and as a function of deep or shallow field.
These same $\rmsD$-versus-$\mSB$ curves 
are used in the {\PSNID} analysis to scale the flux \uncs.
The \PSNID\ analysis results in $\NFAKEPSNID$ fakes passing 
the selection criteria in Table~\ref{tb:PSNID_CUTS} (includes all 10 fields), 
and a similar number of SNe~Ia from the MC simulation.

\begin{figure}[h!]   \centering
\epsscale{1.1} 
\plotone{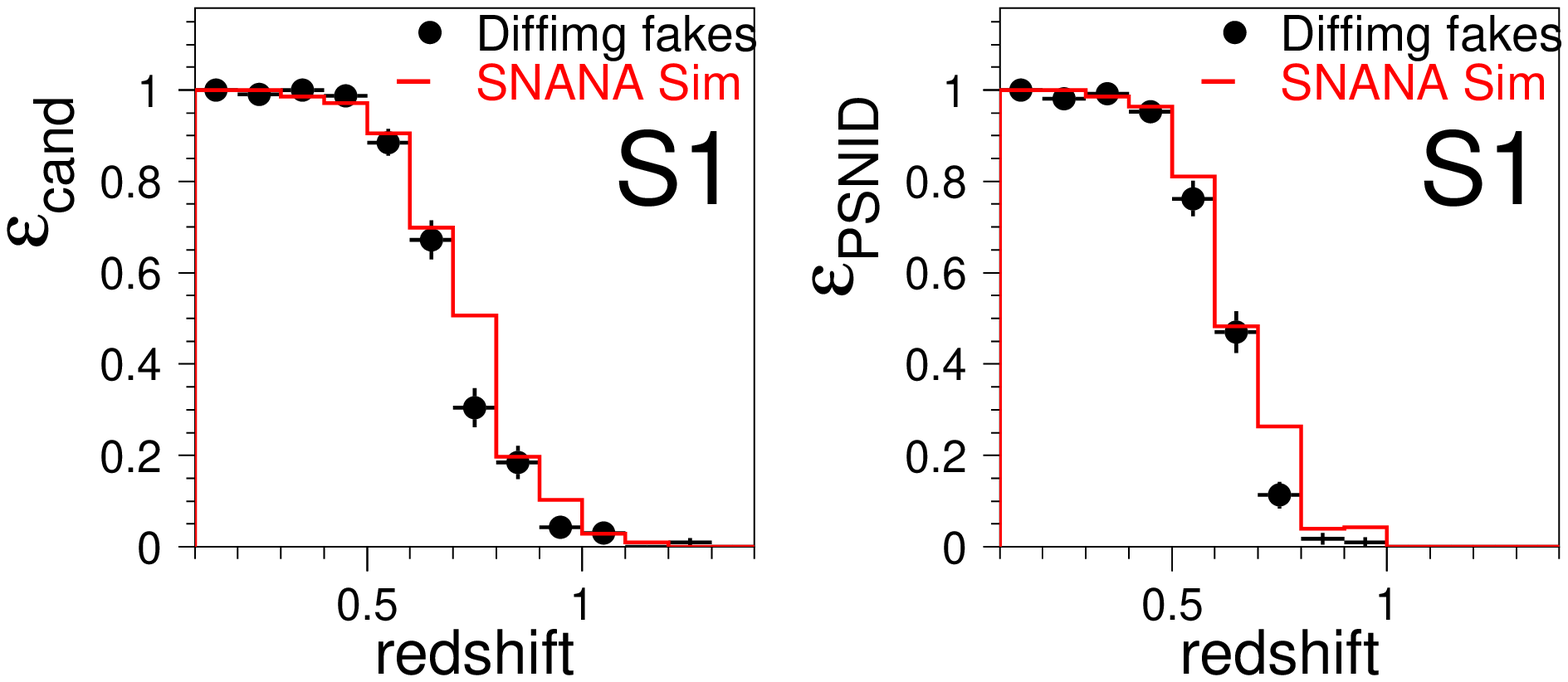}
\plotone{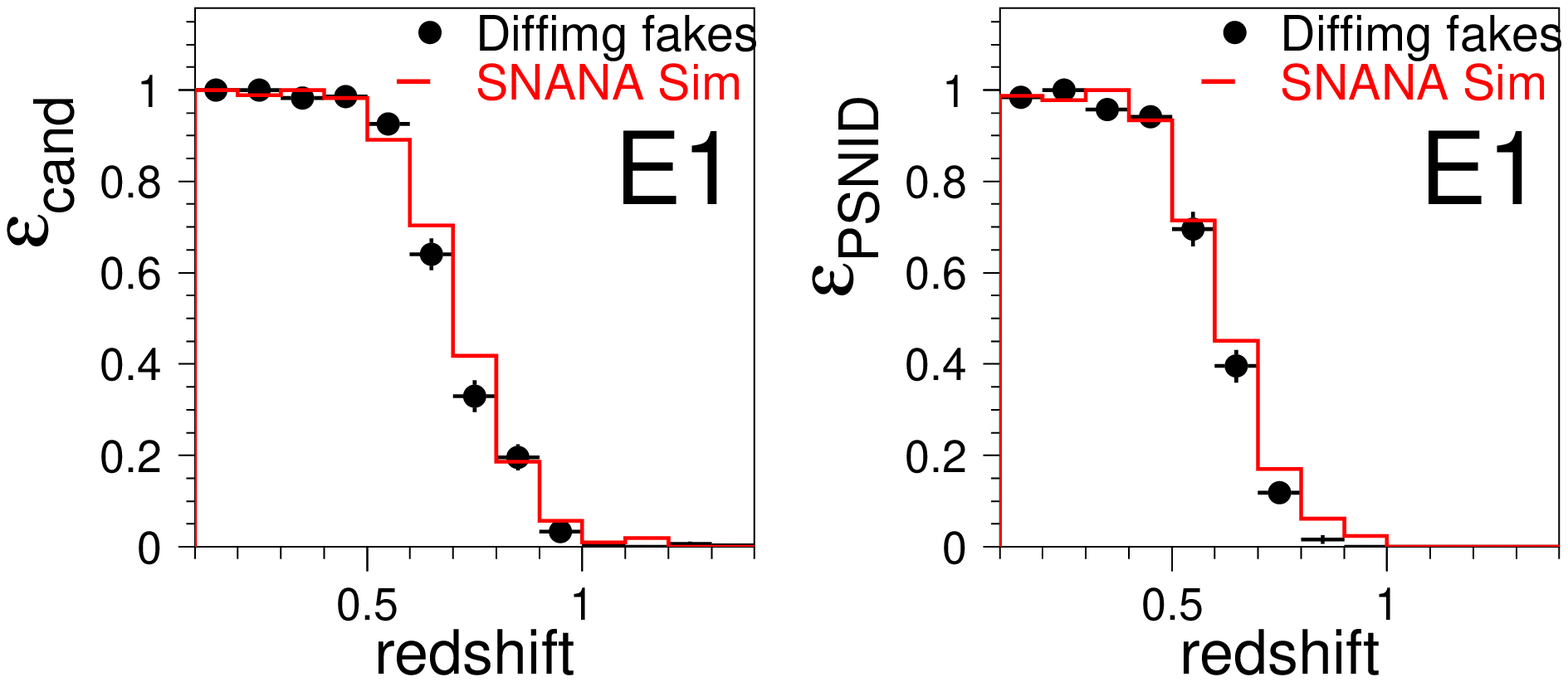}
\plotone{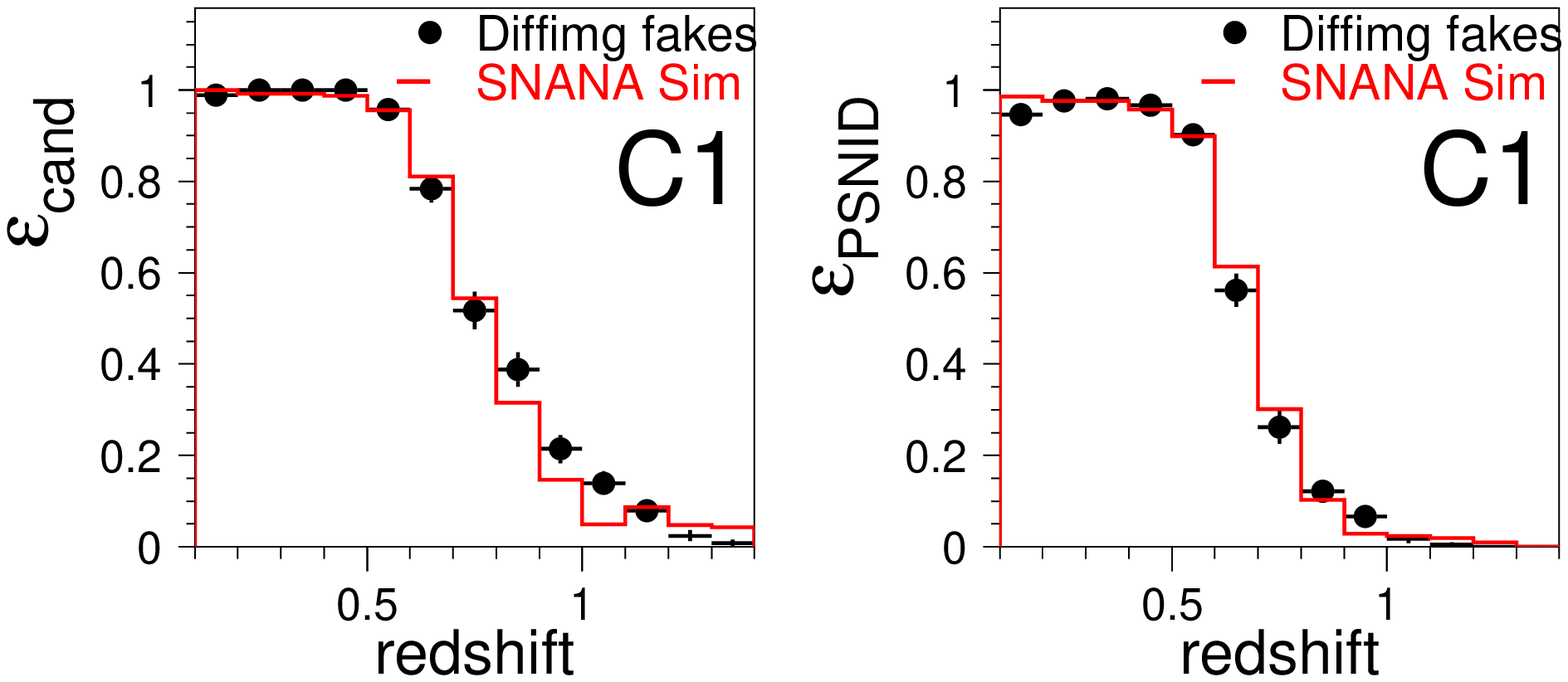}
\plotone{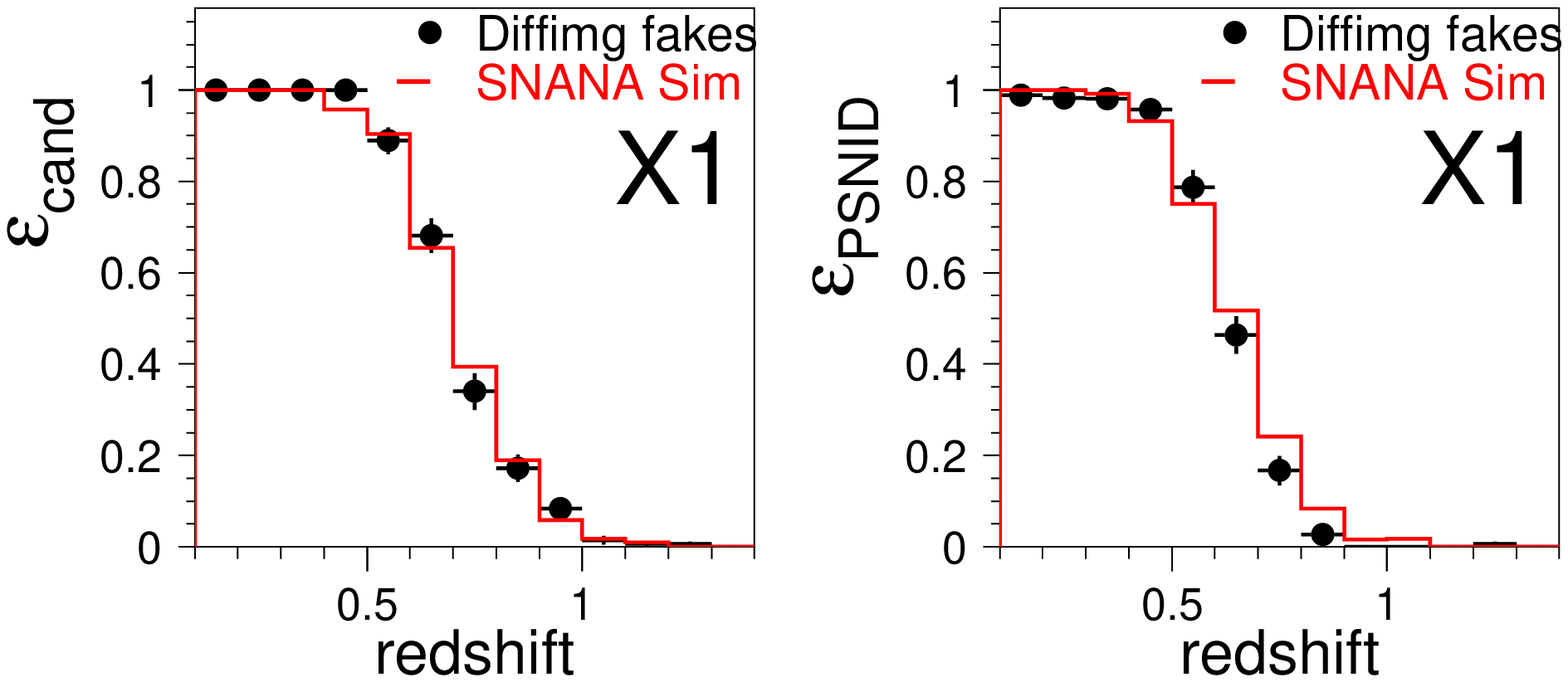}
\caption { 
      For one shallow field in each group, the  Y1 \eff\ vs. redshift is shown for fakes 
      processed by  \Diff\ (black dots), and the MC prediction (red histogram).
      Left panel is the \eff\ for becoming a science \cand\ ($\effCand$);
      right panel is the \PSNID-analysis \eff\ ($\effPSNID$) 
      defined in Table~\ref{tb:PSNID_CUTS}.
	}
  \label{fig:fake_effz_shallow}
\end{figure}

\begin{figure}[h!]   
\centering
\epsscale{1.1} 
\plotone{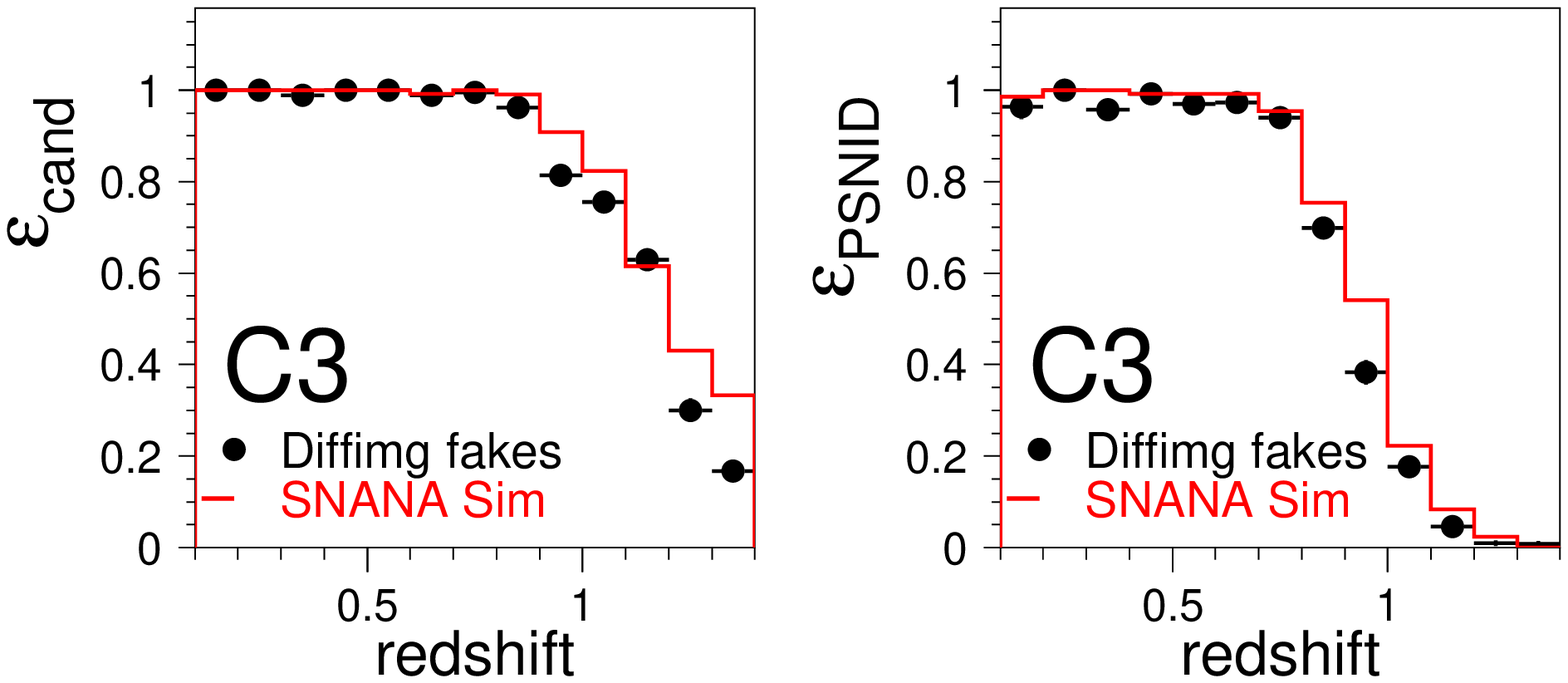}
\plotone{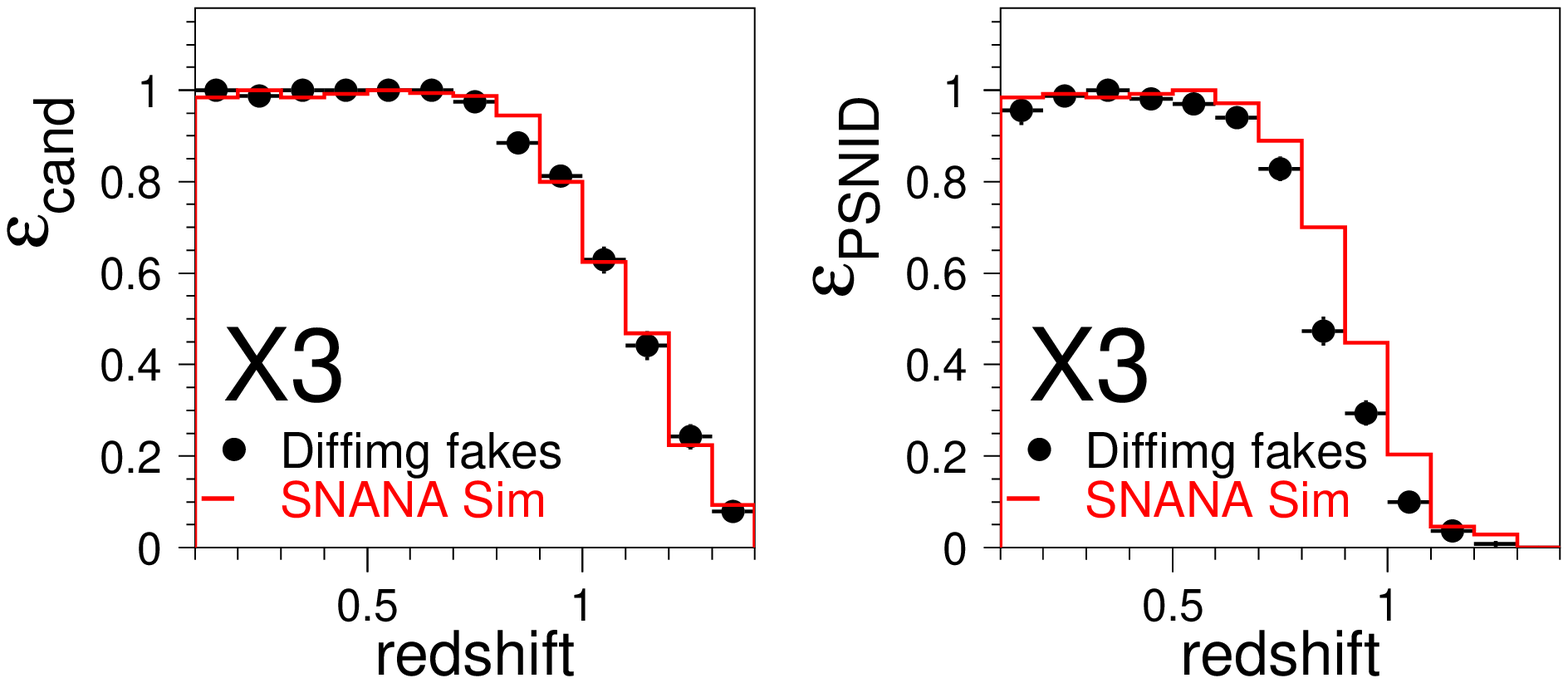}
\caption { 
      Same as Fig.~\ref{fig:fake_effz_shallow}, but for the two deep fields.
	}
  \label{fig:fake_effz_deep}
\end{figure}

Figure~\ref{fig:fake_effz_shallow} shows the science-\cand\   \eff\   ($\effCand$) 
and PSNID-analysis \eff\ ($\effPSNID$) as a function of redshift for one shallow 
field in each group. 
The analogous deep-field plots are shown in Fig.~\ref{fig:fake_effz_deep}.
In the shallow fields, $\effCand \simeq 1$ for redshifts $z<0.5$,
and falls to 50\% at $z\simeq \zShallowEffHalf$. 
In the deep fields, $\effCand \simeq 1$ for redshifts $z<0.8$,
and falls to 50\% at $z\simeq \zDeepEffHalf$. 
The overall agreement is good between the fakes and the MC simulation.
While we might have expected the \SBa\ to affect the discovery of lower 
redshift SNe that preferentialy lie on brighter galaxies, 
we find that the low-redshift \effs\ are $\sim 100$\% and thus the \SBa\ has a 
negligible impact on discovering SNe~Ia. The \SBa\ and its impact are discussed
further in \S\ref{subsec:discuss_SBa}.
The most notable discrepancy is in $\effCand$ for redshifts $z>1.2$ in the C3 deep field,
and  $\effPSNID$ for redshifts $z>0.8$ in both of the deep fields
(Fig.~\ref{fig:fake_effz_deep}).  
Finally, it is worth noting that prior to the final reprocessing the fake \effs\ were 
significantly worse than the MC prediction
for the reasons described in \S\ref{subsec:reproc}.

\subsection{What are the Science Candidates ? } 
\label{subsec:cands}
Here we give a very approximate breakdown for the 
7500 science \cands\ discovered by \Diff\ in Y1,
where each \cand\ requires a \Diff\ detection on 2 separate nights 
with no other selection requirements.
First we use our MC simulation to predict the SN contribution 
(Ia+CC; see Appendix~\ref{app:mc})
and we include events that reach peak brightness well before and after the Y1 season.
We find \NSNCandMC\ SNe, where the \unc\ is from the rate measurements,
and nearly  60\% of the SNe are Type~Ia. 
This SN contribution corresponds to about \candFracSN\% of the \cands.

A non-astrophysical \cand, or artifact, is defined as a \cand\ in which 
more than half of the detections fail the automated scanning requirement 
(\S\ref{ssec:pp} and G15). 
Using this arbitrary but illustrative  definition, $\sim \candFracCRAP$\%
of the science \cands\ are artifacts (i.e., $\sim 2300$ in Y1),
compared with 1.5\% of the fakes.
These artifacts become a science \cand\ because of the relatively loose 
requirement of only 2 detections passing the selection requirements
and automated scanning. 
The relatively small number of artifacts does not cause problems during survey 
operations, and thus we choose to reject them with offline analysis 
software rather than trying to reduce the number of science \cands.

For the remaining science \cands,  a  preliminary assessment of the OzDES  
spectral classifications shows that they are mostly AGN and variable stars.

\bigskip

\section{Reality Check: Data-MC Comparison}
\label{sec:datamc}

Since the \Diff\ results presented so far are  based on fakes and simulations,
here we perform a reality check and compare the \SNANA-based 
MC simulation to Y1 data, where the MC simulation is a mix of 
SNe~Ia and CC SNe as described in  Appendix~\ref{app:mc}.
Recall that the MC simulation has input from fakes processed by  \Diff, 
but there is no tuning with real science \cands.
Here we make a  data-MC comparison for the photometric redshift 
distribution ($\zphot$) of a photometrically selected SN~Ia sample,
using only the SN light curve information.
We do not use any {\specy} confirmed typing information, 
nor do we use any host-galaxy redshifts.

\begin{figure}[h!]   \centering
\epsscale{1.1} 
\plotone{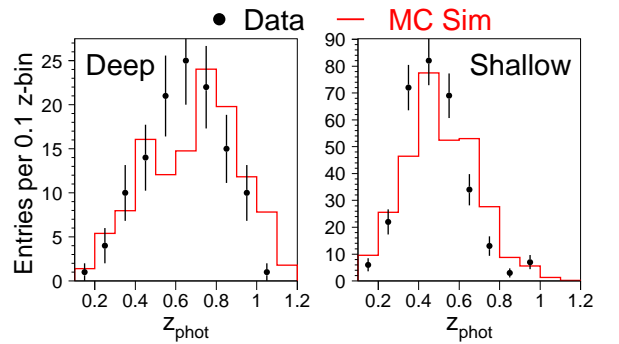}
\caption { 
      $\zphot$ distribution in Y1 from a photometric analysis (Appendix~\ref{app:photana}).
	Left panel is for the 2 deep fields; right is for the 8 shallow fields.
      Black points are the data;  red histogram is the MC prediction,
      re-scaled to have the same number of entries as the data.
	}
  \label{fig:zphot}
\end{figure}

For this comparison we do not use the \PSNID\ selection criteria
in Table~\ref{tb:PSNID_CUTS}.
Instead, we use a more stringent analysis designed to photometrically select a 
highly pure SN~Ia sample.
We fit both the data and MC samples with the \SALTII\  model \citep{Guy2010} 
using the photo-$z$ technique described in \cite{K10_zphot}.
Finally, the \SALTII\ fit parameters are used in a nearest neighbor (NN)
analysis similar to that described in \cite{Sako2014}. 
Details of the analysis are given in Appendix~\ref{app:photana},
and the resulting $\zphot$ comparison is shown in Fig.~\ref{fig:zphot}.
The high-redshift roll-off in the $\zphot$ distributions is mainly
from the requirement that three bands each have an \obs\ with S/N$>5$.
The overall agreement is reasonable, except for $z>1$ in the deep fields.
This data-MC discrepancy will be monitored as we continue
to improve photometric classification methods and the simulation.

\section{Discussion}
\label{sec:discuss}

\subsection{Comparison of Search with SNLS}
\label{subsec:SNLS}

Here we make some rough performance comparisons between the 
SNLS and \DESSN\ deep field search for SNe~Ia.
These two surveys have similar depths and passbands, and they each 
measured their \eff\ with fake SNe~Ia overlaid on images.
While the \DESSN\  trigger requires 2 epochs in any band, 
the SNLS trigger requires a single detection in the Megacam $i_M$ band.
For the single-epoch detection \eff,  Fig.~9 of P10 shows that the magnitude 
at 50\% \eff\ is $\mhalf =  24.3$ in the $i_M$ band for an exposure time of 
3640~s.\footnote{See Table 2 in P10 for SNLS exposure time in each band.}
This depth is very similar to our average DES $i$ band depth, $\mhalf = 24.5$
(Fig.~\ref{fig:mageff50}), using 1440~s exposures.

P10 also measure the \eff\ vs. redshift for finding fake SNe~Ia.
Both the P10 and \SNANA\ simulations predict the observed color 
and stretch distribution for SNLS, and thus the two simulations
are consistent in describing the parent populations of stretch and color.
Fig.~10 of P10 shows that $\effCand=50\%$ at $z\simeq 0.95$,
slightly below the corresponding \DESSN\ redshift  $z\simeq \zDeepEffHalf$.

\subsection{\SBa} 
\label{subsec:discuss_SBa}

As described in \S\ref{sec:monObs},
our image subtractions degrade with increasing galaxy surface brightness,
leading to increased flux scatter 
(see Figsures~\ref{fig:pullSB_deep} and \ref{fig:pullSB_shallow}).
The origin of this \SBa\ has not been identified, but we speculate that it 
may be caused by an underestimate of the pixel flux errors in resampled 
images in the vicinity of bright galaxies.  In particular, resampling 
introduces pixel-to-pixel correlations in the galaxy profile which are not 
included in our estimate of the PSF-fitted \uncs.
Other possibilities include subtle problems in the astrometric solution, 
the PSF determination, or the coadding of exposures.

To check for the possibility that we introduced the \SBa\ in our customized
version of \hotPants, we have run a few tests using the publicly available version.
We find that the subtracted images look very similar, and that our version
results in notably fewer outlier fluxes. 
We are therefore confident that we have not introduced
bugs to cause the \SBa.

In the literature on transient-search pipelines we could not find a
quantitive analysis on the effect of subtractions on bright galaxies.
However, there are some interesting clues in the final-photometry
results reported by \PS\ and SNLS.
In the recent \PS\ cosmology analysis, which uses the same underlying 
subtraction technique as our \Diff, their light curve fits have a reduced 
$\chi^2$ distribution with a larger high-side tail than expected 
(see Fig~6 in \cite{Rest2014}). 
They attribute this effect to subtraction artifacts on bright galaxies,
which is similar to our \SBa.

In the SNLS final photometry \citep{Astier2013}, 
they use a scene modeling technique with stacked images, 
originally developed for SDSS \citep{SMP2008},
which does not use resampled images. 
As a function of total SN + galaxy brightness, they find no evidence for 
flux bias or  scatter (see Figs.~7 and 10 in \citet{Astier2013},
which is encouraging that the \SBa\ can be resolved in the offline analysis.
It is not clear if their lack of \SBa\ is due to a different photometry
method, their astrometric precision being an order of magnitude
better compared to our \Diff,\footnote{
While the SNLS final-photometry pipeline has much better astrometric
precision than our \Diff, the SNLS search pipeline (P10) 
and \Diff\ have similar astrometric precision.} 
or because they do not probe sufficiently bright galaxies to see the effect.

We are actively developing a final-analysis photometry method
similar to that in \citet{SMP2008,Astier2013},
but the \SBa\  may not get resolved for finding transients with \Diff.
The \SBa's impact on  discovering SNe~Ia, however, is quite limited because of 
their brightness at low redshifts where the \SBa\ is most pronounced,
and because only 2 detections are needed among of the many above-threshold \obss.
The main impact is that the larger flux \uncs\ at low redshift slightly 
degrade the classification performance of the \PSNID\ program.

In contrast to bright SNe~Ia, the \SBa\ can have a more dramatic effect on 
detecting and measuring fluxes for faint or fast transients,
such as CC SNe or kilonovae. For example,  kilonova models for neutron-star (NS) 
mergers suggest optical signals that are much dimmer redder, and short-lived
compared to SNe~Ia  \citep{BK2013}. 
Using  \Diff\ to search for such events in very nearby galaxies,
the \SBa\ could significantly degrade the detection \eff,
and  those that are detected could have  color \uncs\ much larger than
expected from photo-electron statistics, thereby making it  difficult to distinguish  
kilonovae from other astrophysical transients.
 
\begin{figure}[h!]   \centering
\epsscale{1.1} 
\plotone{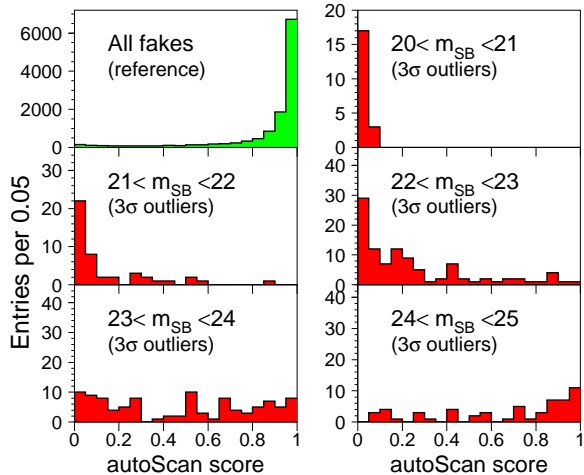}
\caption { 
	\autoScan\ score distribution for fakes in the two deep fields, $i$ band.
      Score is 0 for a cleary bad subtraction and 1 for a good subtraction.
      Upper left panel shows the reference distribution for all fakes.	
	Remaining panels probe the \SBa\  for a subset of fakes
	 that are faint ($m_{\rm SN} >23$),
	are detected by \sex, satisfy the cuts in Table~\ref{tb:obj_cuts},
	and have a measured flux more than $3\sigma$ away from the true flux.
	Each panel indicates a different $\mSB$ range. 
	}
  \label{fig:mlscore}
\end{figure}

To further diagnose the \SBa, Fig.~\ref{fig:mlscore} shows the
\autoScan\ score distribution for $i$ band fake detections in the
two deep fields (X3,C3).   
{\tt AutoScan} assigns a score near zero to a clear artifact, 
and a score near one to a cleanly subtracted point-source transient;
scores above 0.5 are used to make \cands.
The upper-left panel in Fig.~\ref{fig:mlscore} shows the \autoScan\ score
distribution for all of the $i$ band detections; this reference distribution is 
strongly peaked near one, showing that most of the detections are from 
good subtractions.
The remaining panels show the \autoScan\ score distribution,
in bins of $\mSB$, for the small subset of $>3\sigma$ flux outliers.
For the brightest SB range ($20<\mSB<21$) the \autoScan\ scores are all  
close to zero, indicating that these are visibly poor subtractions. 
As the SB decreases, the \autoScan\ scores improve.
We have checked the distributions of PSF, sky noise and ZP,
and find no significant difference between the outliers and the reference;
hence there is no apparent correlation of the \SBa\ with observing conditions.

For \PSNID\ light curve fitting we could remove the few \obss\ that fail \autoScan\
but we do not currently have the infrastructure to apply this requirement to the 
many non-detections that are often more numerous than the  \autoScan\ failures.
As described in \S\ref{sec:monCand},
we have chosen instead to model the  increased flux scatter
and inflate the flux \uncs\ based on $\mSB$.

Finally, we note that our characterization of the bright-galaxy subtraction artifact
is a dependence on a single parameter: $\mSB$. 
While  this description is adequate to classify newly discovered
SNe for \spec\ \obss,  a more accurate
description may be needed for dimmer transients (e.g., kilonovae),
and the Hubble-diagram analysis if this effect persists in the final photometry. 
For example, the \SBa\  could also depend on the exposure conditions
and the SB gradient at the SN location.

\subsection{Host-galaxy Matching} 
\label{subsec:hostMatch}

The SN science analyses will rely mainly on photometric classification, 
and the redshifts will come from host galaxy spectroscopy,
primarily from OzDES \citep{OzDES}. 
The \spec\ redshifts are very accurate in principle, 
if the correct galaxy is matched to each SN.
We have used fakes to measure the SN-host matching performance in Y1,
and found a 99\% success rate.
However, our fakes are preferentially distributed close to the galaxy cores
with too few events in the disk tails, and thus the SN-host matching result
from fakes is too optimistic.

We are therefore preparing to test SN-host matching with an 
independent set of fake locations based on more realistic 
galaxy profiles from semi-analytic models that are fit to S\'ersic profiles.
As we obtain more accurate DES galaxy profiles in 
future analyses, we will be able to use our own data
to evaluate the SN-host matching \eff.  
Also note that fake locations can be rapidly generated and analyzed
at the catalog level since there is no need to overlay SN fluxes on  
images to process with \Diff.
The eventual goal is to update the simulation to include a model of
outlier redshifts from mis-matched host galaxies.

\bigskip \bigskip
\section{Conclusions}
\label{sec:fin}

We have assembled a pipeline capable of using hundreds of CPU cores to process
up to  \MAXDATAVOL~GB of raw imaging data in less than a day,
with the goal of discovering astrophysical transients.
For the subtracted images produced by  \Diff\ in Y1,
the typical number of \sex-detected objects 
per band is a few hundred thousand per 3~\degsq\ field,
and the vast majority ($>90$\%) are subtraction artifacts.
Selection requirements and automated scanning reduce
the artifact fraction down to 25\%, and $\sim 10^4$
detections per band (Table~\ref{tb:obj_stats}). 
The number of detections per single-epoch visit
is $\sim \NDETECT$ per \degsq.

The number of science \cands, requiring a detection on 2 separate nights, 
is \NcandPerDeep\ per deep field, and \NcandPerShallow\ per shallow field
(Table~\ref{tb:cand_stats}). 
Our MC simulation predicts that roughly \candFracSN\% of the 
discovered transients are SNe~Ia or CC SNe.
Another $\sim 30\%$ are artifacts, and most of the remaining \cands\
are AGN or variable stars.

We have implemented extensive monitoring in  \Diff\ based on overlaying
fake SNe~Ia near galaxies on the search images.
Comparing the \Diff\ \eff\ for fakes to the \eff\ from MC simulations shows 
that the \Diff\ performance is close to what is expected.  
The main defect of  \Diff\ is the \SBa\ in which larger 
host-galaxy surface brightness
results in larger flux-scatter that is not described by the \unc\
(see Figures~\ref{fig:pullSB_deep} and \ref{fig:pullSB_shallow}).
There are other small fake-MC discrepancies in the \eff\ 
(e.g.,  Fig.~\ref{fig:fake_effz_deep}); 
it is not clear if the cause is a more subtle \Diff\ defect,
or if the MC simulation is too optimistic.

As a rigorous demonstration of our monitoring technique, 
we performed a very preliminary photometric classification analysis
on real (non-fake) data, and compared the resulting
$\zphot$ distribution to a MC simulation. 
Inputs to the MC simulation include observed conditions 
(PSF, ZP, sky noise)
and the \Diff\ behavior measured with fakes 
(\eff\ vs. S/N and anomalous flux scatter vs. SB). 
The resulting data-MC agreement is reasonable in both the deep and shallow fields
(Fig.~\ref{fig:zphot}).

Finally, the results presented here are based on fully reprocessed data after the first 
two DES seasons. \Diff\ issues during Y1 and Y2 resulted in some poor subtractions, 
but  with  recent \Diff\ improvements and a reliable model of the flux \uncs,
we expect our \spec\ target selection to be  more efficient and 
more automated in the remaining seasons.

\bigskip
\centerline{ACKNOWLEDGMENTS}

This research used resources of the National Energy Research Scientific Computing Center (NERSC), 
a DOE Office of Science User Facility supported by the Office of Science of the 
U.S. Department of Energy under Contract No. DE-AC02-05CH11231.
Part of this research was conducted by the Australian Research Council
Centre of Excellence for All-sky Astrophysics (CAASTRO), through project
number CE110001020.

Funding for the DES Projects has been provided by 
the U.S. Department of Energy, 
the U.S. National Science Foundation, the Ministry of Science and Education of Spain, 
the Science and Technology Facilities Council of the United Kingdom, 
the Higher Education Funding Council for England, the National Center for Supercomputing 
Applications at the University of Illinois at Urbana-Champaign, 
the Kavli Institute of Cosmological Physics at the University of Chicago, 
the Center for Cosmology and Astro-Particle Physics at the Ohio State University,
the Mitchell Institute for Fundamental Physics and Astronomy at Texas A\&M University, 
Financiadora de Estudos e Projetos, 
Funda{\c c}{\~a}o Carlos Chagas Filho de Amparo {\`a} Pesquisa do Estado do Rio de Janeiro, 
Conselho Nacional de Desenvolvimento Cient{\'i}fico e Tecnol{\'o}gico and 
the Minist{\'e}rio da Ci{\^e}ncia, Tecnologia e Inova{\c c}{\~a}o, 
the Deutsche Forschungsgemeinschaft and the Collaborating Institutions in the Dark Energy Survey. 
The DES data management system is supported by the National Science Foundation under Grant Number AST-1138766.
The DES participants from Spanish institutions are partially supported by MINECO under grants AYA2012-39559, ESP2013-48274, FPA2013-47986, and Centro de Excelencia Severo Ochoa SEV-2012-0234, 
some of which include ERDF funds from the European Union.

The Collaborating Institutions are 
Argonne National Laboratory, 
the University of California at Santa Cruz, 
the University of Cambridge, Centro de Investigaciones En{\'e}rgeticas, 
Medioambientales y Tecnol{\'o}gicas-Madrid, 
the University of Chicago, 
University College London, the DES-Brazil Consortium, the University of Edinburgh, 
the Eidgen{\"o}ssische Technische Hochschule (ETH) Z{\"u}rich, 
Fermi National Accelerator Laboratory, 
the University of Illinois at Urbana-Champaign, 
the Institut de Ci{\`e}ncies de l'Espai (IEEC/CSIC), 
the Institut de F{\'i}sica d'Altes Energies, 
Lawrence Berkeley National Laboratory, 
the Ludwig-Maximilians Universit{\"a}t M{\"u}nchen and the associated Excellence Cluster Universe, 
the University of Michigan, the National Optical Astronomy Observatory, 
the University of Nottingham, 
The Ohio State University, the University of Pennsylvania, 
the University of Portsmouth, 
SLAC National Accelerator Laboratory, Stanford University, 
the University of Sussex, and 
Texas A\&M University.

We are grateful for the extraordinary contributions of our CTIO colleagues and the DECam Construction, 
Commissioning and Science Verification teams in achieving the excellent instrument and telescope 
conditions that have made this work possible.  The success of this project also relies critically on the 
expertise and dedication of the DES Data Management group.

\appendix

\section{MC Simulation to Predict the \Diff\ Efficiency}
\label{app:mc}

The fast MC simulation of SNe~Ia is from \SNANA\ \citep{SNANA}.
It uses the exact same generation parameters as those used to generate the fakes
(\S\ref{sss:fakes}):
parent populations of color and stretch, 
intrinsic scatter model,
and a random galaxy location in proportion to its surface brightness density.
For studies requiring SN~CC we use the \SNANA\ simulation 
as described in \citep{K10_SNPCC}.
For studies requiring the absolute rate,
we use the SN~Ia volumetric rate from \cite{Dilday2008}
and the CC rate from \cite{Bazin2009}.
Each simulated epoch corresponds to a real  \obs\ 
in the survey where the model magnitude is converted to an equivalent
forced-photometry flux  using the measured ZP.
The observed PSF and sky noise at each epoch are used to predict the 
measurement \unc, 
\begin{eqnarray}
  \sigSIM^2 = [ F & + & (A\cdot b \cdot \rmsD^2) ] 
  \label{eq:sigSIM}
\end{eqnarray}
where 
$\sigSIM$ is the uncertainty in photoelectrons,
$F$ is the flux,
$A = [2\pi\int {\rm PSF}^2(r,\theta) r dr]^{-1}$ 
is the noise-equivalent area, and
$b$ is the effective sky level including dark current, readout noise,
and noise from the host galaxy.
$\rmsD$ is an empirical error scaling of the sky noise that increases with the 
local surface brightness as shown in 
Figures~\ref{fig:pullSB_deep} and \ref{fig:pullSB_shallow};
this term accounts for the \SBa: systematic subtraction problems near bright galaxies.
While the measured $\rmsD$ curves are used to compute 
anomalous  fluctuations in the measured fluxes, 
the reported \uncs\ are computed with $\rmsD=1$
in the same way as the data. 

The simulation  includes the \cand\ selection requirement of a detection 
on two separate nights. The detection \eff\ is computed from the $\effSNR$ 
curves in Fig.~\ref{fig:fake_effSNR}. A simulated detection requires
$\effSNR > r$, where $0<r<1$  is a random number.

Finally, the simulated light curves are stored in data files and analyzed 
in exactly the same way as transients (fakes or real events) found by \Diff.

\section{Photometric Analysis and Selection Requirements}
\label{app:photana}

Here we describe a photometric analysis and selection requirements 
to obtain a high-purity sample of SNe~Ia in the first season of the DES-SN program.
The goal of this analysis is to compare the $\zphot$ distribution for data and the
MC simulation.
Using the \SALTII\ model, light curve fitting is done with
the \SNANA\ program {\tt snlc\_fit.exe}.  For each candidate,
the 5 fitted parameters  are
(1) time of peak brightness ($t_0$), 
(2) \SALTII\ color parameter ($c$), 
(3) \SALTII\ stretch parameter ($x_1$), 
(4) \SALTII\ amplitude ($x_0$), and
(5) photometric redshift ($\zphot$).

The first fit iteration chi-squared ($\chi_1^2$) is computed in the usual manner:
from the data-model flux-difference for each epoch, 
and the quadrature sum of the data and model \uncs. 
Since the model \unc\ depends on the fitted parameter $\zphot$, 
the second fit iteration chi-squared ($\chi_2^2$) is
\begin{equation}
    \chi_2^2 = \chi_1^2 + \chisqSigma ~~~~{\rm where} ~~~~
   \chisqSigma = \sum_e 2\ln(\sigma^e/\sigma^e_1)~.
   \label{eq:chisqSigma}
\end{equation}
The index $e$ is the epoch index and $\sigma^e_1$ is the quadrature 
sum of the data and model-\unc\  from the first fit iteration in which 
there is no $\chisqSigma$ term. While the $\sigma^e_1$ add an 
irrelevant constant to $\chi_2^2$, it has the effect of making 
$\chisqSigma$ small. The analysis selection requirements are as follows:
\begin{enumerate}
   \item three bands with at least one \obs\ satisfying S/N$>5$.
   \item at least 1 \obs\ with $\Trest < -2$~days, where $\Trest \equiv \Tobs/(1+\zphot)$.
   \item at least 1 \obs\ with $\Trest > +10$~days.
   \item  \SALTII\ stretch parameter $|x_1| < 4$
   \item $0.02 < \zphot < 2$
   \item fit probability $\Pfit > 0.1$, calculated from fit  $\chi^2$/dof. 
   \item $|\chisqSigma| < 2.5$.
   \item NN requirement described below.
\end{enumerate}

\newcommand{\Dz}{\Delta_z}
\newcommand{\Dc}{\Delta_c}
\newcommand{\Dx}{\Delta_{x_1}}
\newcommand{\DB}{\Delta_B}

The  NN analysis is based on the four-dimensional space
of $x_1$, $c$, $\zphot$ and $\restB$. The first three variables are from the
\SALTII\ light curve fit (see above).
$\restB$ is the true rest-frame $B$-band magnitude as described in  Sec 4.3 of \cite{K13},
and is not the naive best-fit model magnitude. 
For a given set of fitted parameters, the  NNs are simulated events 
that satisfy a four-dimensional distance constraint,
\begin{equation}
    d^2 =  \left[
    \frac{(c-c')^2}{\Dc^2}  + \frac{(x_1-x_1')^2}{\Dx^2} + 
    \frac{(\zphot-\zphot')^2}{\Dz^2} + \frac{(\restB-\restB')^2}{\DB^2}   \right]
     < 1
     \label{eq:NNdist}
\end{equation}
where the primed quantities are the fitted parameters from a simulated
training sample that includes SNe~Ia and CC SNe events.
The optimal distance-metric parameters ($\Dc,\Dx,\Dz,\DB$) are trained
with the simulation to maximize the product of the SN~Ia purity and the \eff.
The final selection requirement is that for simulated neighbors satisfying Eq.~\ref{eq:NNdist},
more than half are true SNe~Ia with at least $1\sigma$ confidence.

\bigskip
\bibliographystyle{apj}
\bibliography{DiffImg_ms}  

  \end{document}